\documentclass[onecolumn]{IEEEtran}
\usepackage{amssymb,amsmath,amsfonts,amsthm}
\usepackage{color}
\usepackage[sort,nocompress]{cite}

\DeclareMathOperator*{\argmax}{arg\,max}
\usepackage{mathtools}
\usepackage{float}
\usepackage{mathrsfs}
\usepackage{relsize}
\usepackage{graphicx}		
\usepackage[font={small}]{caption}
\restylefloat{table}
\usepackage{tabularx}
\usepackage{booktabs}
\usepackage{enumerate}
\theoremstyle{plain}
\usepackage{units}
\usepackage[numbers]{natbib}
\usepackage{bm}
\usepackage{algorithm}

\newtheorem{lem}{Lemma}

\theoremstyle{definition}
\newtheorem*{mydefproof*}{Proof}

\newcommand*\diff{\mathop{}\!\mathrm{d}}
\allowdisplaybreaks

\begin{document}

\title{Stochastic Geometric Analysis of\\ Energy-Efficient Dense Cellular Networks}
\author{Arman Shojaeifard, \IEEEmembership{Member,~IEEE}, Kai-Kit Wong, \IEEEmembership{Fellow,~IEEE},\\ Khairi Ashour Hamdi, \IEEEmembership{Senior Member,~IEEE}, Emad Alsusa, \IEEEmembership{Senior Member,~IEEE},\\ Daniel K. C. So, \IEEEmembership{Senior Member,~IEEE}, and Jie Tang, \IEEEmembership{Member,~IEEE}
\thanks{\scriptsize{A. Shojaeifard and K.-K Wong are with the Department of Electronic and Electrical Engineering, University College London, London, UK (e-mail: a.shojaeifard@ucl.ac.uk; kai-kit.wong@ucl.ac.uk). K. A. Hamdi, E. Alsusa, and D. K. C. So are with the School of Electrical and Electronic Engineering, University of Manchester, Manchester, UK (e-mail: k.hamdi@manchester.ac.uk; e.alsusa@manchester.ac.uk; d.so@manchester.ac.uk). J. Tang is with the School of Electronic and Information Engineering, South China University of Technology, Guangzhou, China (e-mail: eejtang@scut.edu.cn). \par This work was supported by the UK Engineering and Physical Sciences Research Council (EPSRC) under grants EP/N008219/1 and EP/J021768/1. \par This paper was presented in part at the IEEE Vehicular Technology Conference (VTC Spring), Nanjing,
China, May 2016. \cite{7504384}.}}
}

\maketitle

\begin{abstract}
Dense cellular networks (DenseNets) are fast becoming a reality 
with the rapid deployment of base stations (BSs) aimed at meeting the explosive data traffic demand. In legacy systems however this comes with the penalties of higher network interference and energy consumption. In order to support network densification in a sustainable manner, the system behavior should be made `load-proportional' thus allowing certain portions of the network to activate on-demand. In this work, we develop an analytical framework using tools from stochastic geometry theory for the performance analysis of DenseNets where load-awareness is explicitly embedded in the design. The model leverages on a flexible cellular network architecture where there is a complete separation of the data and signaling communication functionalities. Using the proposed model, we identify the most energy-efficient deployment solution for meeting certain minimum service criteria and analyze the corresponding power savings through dynamic sleep modes. Based on state-of-the-art system parameters, a homogeneous pico deployment for the data plane with a separate layer of signaling macro-cells is revealed to be the most energy-efficient solution in future dense urban environments.
\end{abstract}


\section{Introduction}

Ultra-dense deployment of base stations (BSs), relay nodes, and distributed antennas is considered a de facto solution for realizing the significant performance improvements needed to accommodate the overwhelming future mobile traffic demand \cite{6736747}. While legacy wireless communication systems are fast approaching the information-theoretic capacity limits, dense cellular networks (DenseNets) can push data rates even further by shortening the transmitter-receiver distance and serving fewer users per cell \cite{6477048}. The extremely populated topology of DenseNets raises several technical challenges, including managing the aggregate network interference and keeping the energy expenditure in check - the main topics of this paper.    

Understanding the interference behavior in DenseNets is challenging due to the rapid, irregular, and overlapping placement of nodes. In addition, in contrast to existing macro-cells where different parts of spectrum is allocated to neighboring cells, DenseNets employ an aggressive frequency reuse strategy where different nodes can access the same spectrum; thus highlighting the importance of interference management for facilitating efficient spectrum utilization \cite{7118688}. On the other hand, legacy cellular networks and transmission technologies are designed and dimensioned to meet the coverage and capacity requirements in peak traffic conditions. This approach threatens the commercial viability of deploying many more network nodes which would substantially increase the total capital, operational, and environmental expenditure \cite{5978416}, \cite{7177127}. An extensive design overhaul towards a flexible cellular network architecture with load-proportional energy consumption behavior is hence needed. 


In recent years, several collaborative initiatives, such as the GreenTouch consortium \cite{101}, have focused on analyzing the achievable spectral and energy efficiency performance of network densification as well as other promising solutions, using mostly system-level simulations. This approach is inline with the traditional system planning where Monte-Carlo simulations are utilized for drawing conclusions on the cellular network performance. However, due to the inherent characteristics of DenseNets, the simulation-based investigations have become extremely resource-intensive. In order to reduce the underlying complexity associated with DenseNet planning, tractable and computationally-efficient analytical models are deemed necessary for depicting the fundamental bounds and trade-offs. Tools from applied probability theory, in particular stochastic geometry and point processes, are well-suited for characterizing the key performance metrics of DenseNets with random topologies, see \cite{6171996}, \cite{6126035}, and \cite{6365639}.

Despite insightful efforts, however, the common set of assumptions for studying cellular networks using stochastic geometry theory are benign, since rigorous analysis based on a direct signal-to-interference-plus-noise ratio (SINR) probability density function (pdf) approach is challenging \cite{6516171}, \cite{7277029}. Most existing works thus resort to a limiting Rayleigh fading channel model which allows for the exact derivation of the SINR pdf \cite{6524460}, \cite{5223626}. In \cite{6516171}, the authors utilize the non-direct moment-generating-function (MGF) methodology from our previous work in \cite{5407601} to characterize the average rate of always-full-buffer multi-tier cellular networks over arbitrary fading interference channels. However, neglecting network load by assuming that every deployed BS is always-transmitting leads to an unrealistic fully-loaded interference field which severely limits the achievable gains jointly in terms of data rate, deployment cost, and energy efficiency. 

The work in \cite{6463498} shows that a significant gain in coverage probability performance can be achieved through conditional thinning of the interference field as a function of users' density. Moreover, the work in \cite{6575091}, with a focus on energy efficiency, computes the optimal deployment density in load-dependent two-tier cellular networks by numerically fitting the Poisson-Voronoi cell sizes using the Gamma function. Besides the lack of a formulation for the exact pdf of cell sizes, in \cite{6575091}, the impact of SNR operating regions and artificial-bias on achievable performance were not investigated. 
Furthermore, the dynamic switching of nodes in \cite{6575091} is limited due to the global coverage constraint required by the inflexible traditional cellular network architecture where BSs must simultaneously provide coverage and capacity.    


Recently, a design overhaul in the cellular network architecture with a complete separation of the control and data infrastructures has been proposed through the Beyond Green Cellular Green Generation (BCG$^{\text{2}}$) project within GreenTouch consortium \cite{6152217}, \cite{6533307}. The control network is responsible for providing continuous global coverage so that communication services can be requested by users at any given time and location. A sparse overlay of large signaling-only cells with extended range is considered to be the best solution in terms of deployment and energy costs for this functionality. The data network, on the other hand, is in charge of delivering communication services by intelligently activating resources on-demand based on an optimal selection of the access device which can meet the service requirements at minimum energy cost. The network layout for the capacity plane can in general be heterogeneous but pinpointing the optimal deployment solution remains an open problem. 

In \cite{6918448}, we incorporated the notion of inherent spatial-correlations and load-proportionality in the design and analysis of heterogeneous networks (HetNets). A computationally-efficient stochastic geometry-based framework for calculating the average rate of a typical user in the HetNet paradigm was accordingly provided. It was shown that the average rate performance of realistic correlated load-aware HetNets is significantly more optimistic that the state-of-the-art results which consider BSs to be either always or independently transmitting. In this paper, we extend the work in \cite{6918448} to identify the most energy-efficient combination of BS densities and the corresponding power savings through dynamic BS switching in multi-tier cellular networks. It should be noted that stochastic geometry theory cannot provide insight into the precise locations of network sources. The proposed methodology can however give us valuable information concerning the optimal type and number of BSs needed to satisfy certain user requirements. This approach can also be used to identify how many BSs can be switched off for the purpose of energy savings under fluctuations in the traffic volume.  


\begin{figure}[t]
\centering
\includegraphics{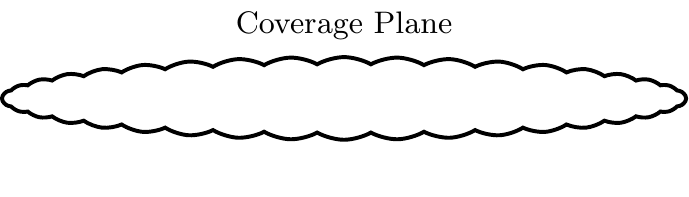}
\vspace*{-2em}
\end{figure}
\begin{figure}[t]
\centering
\includegraphics{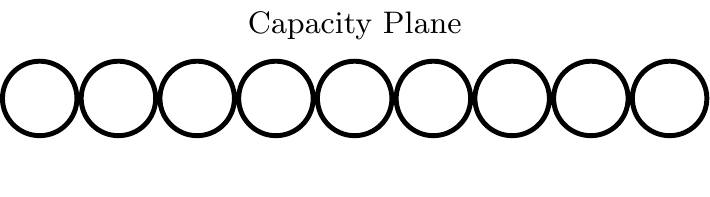}
\caption{Green cellular architecture based on the separation of the control and data networks.}
\label{GreenArch}
\end{figure}

\section{Paper Contribution}

Here, we focus on the green cellular network architecture in Fig. \ref{GreenArch}, where global coverage is provided by sparse large signaling-only cells and capacity is injected on-demand using dense data-only BSs. This allows for greater flexibility in utilizing sleep modes in the capacity plane which is no longer constrained by the global coverage constraint. Considering randomly-deployed cellular networks, we incorporate the notion of load-proportionality and correlated interfering sources by optimally and exclusively associating every user equipment (UE) to a data-only BS which provides the greatest reward under arbitrary shadowing characteristics. Closed-form expressions for the statistics of the received signal power and aggregate network interference over Nakagami-m fading channels are accordingly provided towards efficient computation of the average rate. An optimization problem for computing the optimal deployment solution that minimizes the total energy expenditure whilst satisfying a minimum rate requirement under a given network load is hence formulated and tackled using exhaustive search algorithms. For the special case of homogeneous interference-limited DenseNets, we provide new closed-form bounded solutions of the average rate and optimal BS density. Strategic sleep modes are then utilized for realizing power savings according to the temporal fluctuations in the traffic volume. The validity of the proposed analytical framework and its advantages in terms of preserving energy over the state-of-the-art fully-loaded and interference-thinning-based models are demonstrated via Monte-Carlo trials.

Several useful design guidelines are concluded from our findings. In general, we show that the minimum deployment density needed to satisfy the traffic requirements is significantly smaller in realistic load-proportional cellular networks than existing results which typically assume independence among the BS activities. Furthermore, we show that on the contrary to the fully-loaded and interference-thinning approaches, our analytical model closely matches the actual optimal deployment density. The implications of these trends on the overall energy consumption of the network are accordingly highlighted. In addition, the results confirm the promising potential of network densification towards effective offloading of traffic from large-cells. Artificial-expansion of small-cells coverage range, on the other hand, is shown to only further improve performance under low network loads. We further depict the limitations of large-cells in interference-dominant operating regions; a similar trend is observed for small-cells in noise-dominant regions. Under anticipated traffic and rate requirements for dense urban environments in the year 2020, the optimal deployment solution for the capacity plane is calculated to be a homogeneous pico network, capable of realizing peak power savings of near 15 kW/km$^{\text{2}}$ over a conventional macro-only data network. 

\section{Paper Organization}

The reminder of the paper is organized as follows. The system model and mathematical preliminaries are described in Section \ref{secModel}. In Section \ref{rateSec}, an analytical framework for computationally-efficient calculation of the average rate is provided. An optimization problem for identifying the optimal deployment solution is formulated in Section \ref{eeSec}. The power savings analysis with dynamic switching of BSs is discussed in Section \ref{psSec}. In Section \ref{secRES}, theoretical and simulation studies are conducted towards unveiling network design pointers. Finally, the paper is concluded in Section \ref{secCON}.  

\textit{Notation:} $\mathbb{E}_{x}\{.\}$ denotes the expectation operator with respect to random variable $x$; $\mathscr{P}(x)$ is the probability of event $x$; $\mathcal{P}_{x}(.)$ represents the pdf of random variable $x$; $\mathcal{M}_{x}(z) = \mathbb{E}_{x} \left\{ \exp \left( - z x \right) \right\}$ is the MGF of random variable $x$; $| . |$ is the modulus operator; $\| . \|$ corresponds to the Euclidean distance; $\Gamma(x) = \int^{+ \infty}_{0} \exp \left( - s \right) s^{x - 1} \diff s$ is the Gamma function; $\Gamma(y,x) = \int^{+ \infty}_{x} \exp \left( - s\right) s^{y - 1} \diff s$ is the (upper) incomplete Gamma function; $_2F_1(a,b;c;d) = \sum^{+ \infty}_{x = 0} \, \frac{(a)_{x} (b)_{x}}{ (c)_{x} \, x!} d^{x}$, where $(n)_{x} = n (n + 1) ... (n + x -1)$, is the Gauss hypergeometric function.

\section{System Model and Mathematical Preliminaries}
\label{secModel}

Consider the downlink capacity plane comprising UEs and $T$ different classes of load-proportional BSs respectively distributed according to stationary homogeneous Poisson point processes (PPPs) $\Phi^{(u)}$ and $\Phi^{(b)}_{t}$ with spatial densities $\lambda^{(u)}$ and $\lambda^{(b)}_{t}$, where $t \in \mathcal{T} = \{1,2,... T\}$. We assume global coverage is maintained through deploying, or utilizing the already in place, legacy macro-cells; the reader is referred to \cite{6664198} for information on the operational characteristics and energy savings procedures in the separated coverage plane. We consider a co-channel deployment with universal frequency reuse where each operating BS equally allocates resources in terms of time or frequency slots to its associated UEs \cite{6497017}, \cite{6497439}. This implies that there is no interference from transmissions associated with the same BS. For the sake of analytical tractability, we assume all tier-$t$ BSs to have equal transmit power $P^{tx}_{t}$, artificial-biasing weight $\beta_{t}$, and path-loss intensity $\alpha_{t}$. Let $\| Y_{t,l,k} \|$, $H_{t,l,k}$, $\chi_{t,l,k}$, and $P^{rx}_{t,l,k}$ denote the Euclidean distance, fading power gain, shadowing power gain, and received signal power at the $k$-th UE from the $l$-th tier-$t$ BS, respectively. In addition, the constant additive noise power is denoted by $\eta$. Note that the framework can be extended to the case where all nodes are equipped with multiple antennas using the methodology in \cite{7478073}.

We consider a cellular association and load-balancing strategy where every active user connects to the closest BS of a certain tier which provides the strongest shadowed received signal power mathematically formulated as follows
\begin{gather}
\left( t^{*}_{\in \mathcal{T}} , l^{*}_{\in \Phi^{(b)}_{t}} \right) = \argmax \left( \beta_{t} P^{tx}_{t,j,k} \chi_{t,j,k} \| Y_{t,j,k} \|^{- \alpha_{t}} \right) \; , \; \forall t \in \mathcal{T} , \forall j \in \Phi^{(b)}_{t}, \forall k \in \Phi^{(u)} \\ 
\text{subject to:} \;\;\; \varphi_{t,j,k} \in \left\{ 0,1 \right\} \; , \; \forall t \in \mathcal{T}, \forall j \in \Phi^{(b)}_{t}, \forall k \in \Phi^{(u)} \label{cons112}\\
\sum_{t \in \mathcal{T}} \sum_{j \in \Phi^{(b)}_{t}} \varphi_{t,j,k}  = 1 \; , \; \forall k \in \Phi^{(u)} \label{cons122}
\end{gather}
where $\varphi_{t,l,k}$ is a binary decision variable depicting whether or not the $k$-th UE is served by the $l$-th tier-$t$ BS and constraints in (\ref{cons112}) and (\ref{cons122}) ensure that each UE is exclusively associated with exactly one BS. Accordingly, the optimal binary decision variables of all UEs are selected. 

The corresponding SINR of the $k$-th UE served by the $l^{*}$-th tier-$t^{*}$ can hence be expressed as
\begin{gather}
\gamma_{t^{*},l^{*},k} = \frac{P^{rx}_{t^{*},l^{*},k}}{\eta + I_{agg,k}}
\end{gather}
where
\begin{align}
P^{rx}_{t^{*},l^{*},k} = P^{tx}_{t^{*}} H_{t^{*},l^{*},k} \chi_{t^{*},l^{*},k} \| Y_{t^{*},l^{*},k} \|^{- \alpha_{t^{*}}} 
\end{align}
and
\begin{align}
I_{agg,k} = \bigcup_{c \in \Phi^{(u)} / \{ k \}} \varphi_{t,l,c} \sum_{t \in \mathcal{T}} \sum_{l \in \Phi^{(b)}_{t} \backslash \{ l^{*}\} } P^{tx}_{t} H_{t,l,k} \chi_{t,l,k} \| Y_{t,l,k} \|^{- \alpha_{t}}.
\end{align}

Based on the results from \cite{6658810} and considering identical distribution across links, shadowing effects can be interpreted as random displacements in the original BSs locations using new transformed PPPs $\Phi^{(b)}_{t,s}$ with 
\begin{align}
\lambda^{(b)}_{t,s} = \lambda^{(b)}_{t} \mathbb{E} \left\{ \chi^{2/\alpha_{t}}_{t} \right\}
\end{align}
iff $\mathbb{E} \left\{\chi^{2/\alpha_{t}}_{t} \right\} < \infty$, $\forall t \in \mathcal{T}$.  As an application example, we consider Log-Normal shadowing with mean $\mu_{t}$ (dB) and standard deviation $\sigma_{t}$ (dB), where $t \in \mathcal{T}$. 
Note that fading does not impact cell selection as it can be averaged or equalized using narrowband partitioning schemes 
\cite{6567878}. 

In this work, we consider independent unit-mean Nakagami-m fading for intended and interfering links. In this case, the pdf and MGF of the normalized channel power gain between the $l$-th tier-$t$ BS and $k$-th UE are respectively expressed as \cite{6152071}
\begin{align}
& \mathcal{P}_{H_{t,l,k}}(h) = \frac{ m_{t}^{m_{t}} h^{m_{t} - 1} }{\Gamma \left( m_{t} \right)} e^{- m_{t} h}
\end{align}
and
\begin{align}
\mathcal{M}_{H_{t,l,k}}(z) = \left( 1 + \frac{z}{m_{t}} \right)^{-m_{t}}
\label{NakaMGF}
\end{align}
where $m_{t}$, $t \in \mathcal{T}$, is the Nakagami-m fading parameter which can fit a wide-range of wireless channel fading models. 

\begin{figure}[!t]
\centering
\includegraphics{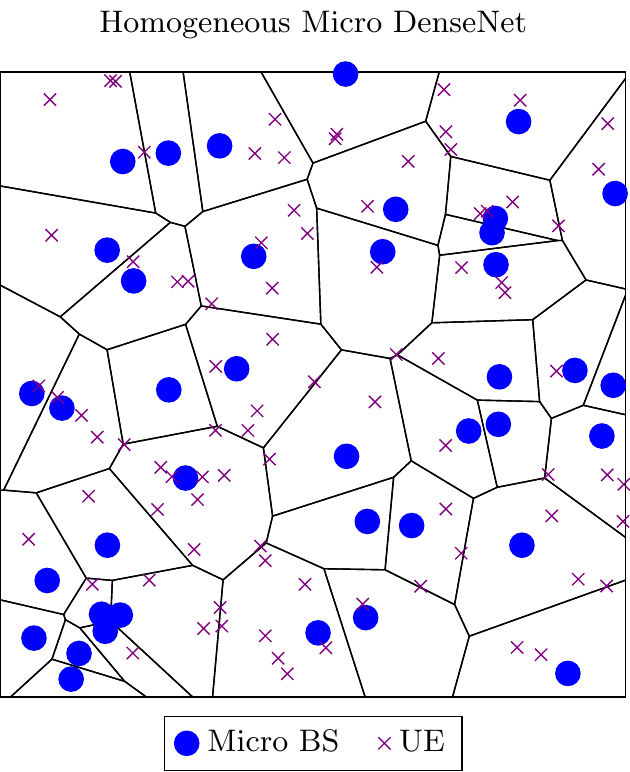}    
\caption{Poisson-Voronoi tesellations in a single-tier micro DenseNet, $5\times5$ $\text{km}^{2}$ area, $\lambda^{(b)}_{m}  =  6$ $\text{BSs/km}^{2}$, $\lambda^{(u)}  =  15$ $\text{UEs/km}^{2}$, $P^{tx}_{m}  =  6.3$ W, $\mu_{m} = 0$ dB, $\sigma^2_{m} = 1$ dB, $\alpha_{m}  =  4$.}
\label{SingleTierDN}
\end{figure}

\begin{figure}[!t]
\centering
\includegraphics{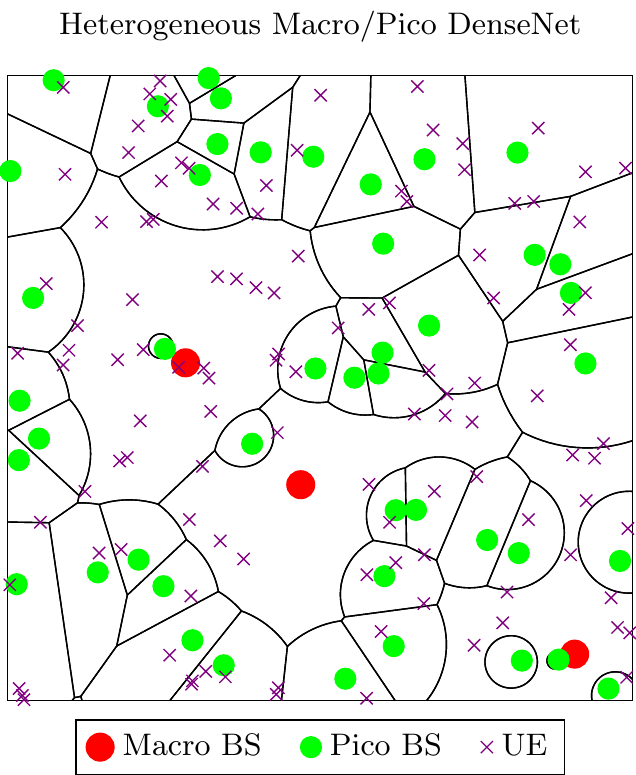}
\caption{Multiplicatively-weighted Poisson-Voronoi tessellations in a two-tier macro/pico DenseNet, $5\times5$ $\text{km}^{2}$ area, $\lambda^{(b)}_{M} = 0.1$ $\text{BSs/km}^{2}$, $\lambda^{(b)}_{p} = 8$ $\text{BSs/km}^{2}$, $\lambda^{(u)}  =  15$ $\text{UEs/km}^{2}$, $P^{tx}_{M} =  20$ W, $P^{tx}_{p} = 0.13$ W, $\beta_{M} = 0$ dB, $\beta_{p}  = 18$ dB, $\mu_{M} = \mu_{p} = 0$ dB, $\sigma^2_{M} = \sigma^2_{p} = 1$ dB, $\alpha_{M} = \alpha_{p} =  4$.}
\label{MultiTierDN}
\end{figure}

The ideal energy consumption behavior of a cellular network is load-proportional where the whole system power usage varies linearly according to the network load, i.e., from operating at maximum power under full-load to almost zero when there is no traffic. The importance of this concept can be depicted using illustrative examples. Consider the topology of a single-tier micro DenseNet in Fig. \ref{SingleTierDN}, where the capacity regions are formed according to the closest BS-UE distances, resulting in a classical Poisson-Voronoi tessellation. Based on the impractical fully-loaded assumption, which is widely used in the literature for the sake of analytical tractability, all deployed nodes are transmitting. It can however be seen from Fig. \ref{SingleTierDN} that even though the UEs density is relatively large, there are certain BSs that are inactive. A similar trend can be observed for the multiplicatively-weighted Poisson-Voronoi tessellation topology of a two-tier macro/pico DenseNet in Fig. \ref{MultiTierDN}. By making the network behavior load-proportional, BSs are only turned on when needed thus substantially enhancing the energy efficiency of the system. Note that the practical feasibility of adopting this approach is facilitated through separating the cellular network signaling and data infrastructures. 



\section{Average Rate Performance}
\label{rateSec}

In this section, we provide a framework for calculating the average communication rate achievable by an arbitrary user in the DenseNet paradigm. The Shannon channel capacity formula, i.e., $\log_{2} \left( 1 + \text{SINR} \right)$ b/s/Hz, is applicable here assuming capacity-achieving codes are used for the operating instantaneous SINR. Note that the model can be easily adjusted to capture other modulation/coding schemes by adding a SINR gap to the instantaneous rate formula, i.e., $\log_{2} \left( 1 + \frac{\text{SINR}}{\Gamma} \right)$ b/s/Hz, where $\Gamma$ ($\geq 1$) denotes the SINR gap. Hence, the average rate in b/s/Hz of an arbitrary UE $k$ assumed to be located at the origin can be mathematically formulated by
\begin{align}
\mathcal{R} \left(\lambda^{(u)},T,\lambda^{(b)}_{t},P^{tx}_{t},\beta_{t},\eta,\alpha_{t},m_{t},\mu_{t},\sigma_{t} \right) = \sum_{t^{*} \in \mathcal{T}} \sum_{l^{*} \in \Phi^{(b)}_{t^{*}}} \overline{\varphi}_{t^{*},l^{*},k} \mathbb{E} \left\{ \log_{2} \left( 1 + \gamma_{t^{*},l^{*},k} \right) \right\}
\label{InsightRate}
\end{align}
where $\overline{\varphi}_{t^{*},l^{*},k}$ is used to denote the probability of UE $k$ being connected to the $l^{*}$-th tier-$t^{*}$ BS and $\mathbb{E} \left\{ \log_{2} \left( 1 + \gamma_{t^{*},l^{*},k} \right) \right\}$ is the average rate of UE $k$ conditioned on its association to the $l^{*}$-th tier-$t^{*}$ BS. 

In \cite{5407601}, \textit{Hamdi} showed that the capacity evaluation of wireless communication systems can be greatly simplified by expressing the averages like $\mathbb{E} \left\{ \log_{2} \left( 1 + \frac{X}{Y + 1} \right) \right\}$ in terms of the MGFs of the independent random variables $X$ and $Y$, i.e., $\int^{+ \infty}_{0} \mathcal{M}_{Y} \left( z \right) \left[ 1 - \mathcal{M}_{X} \left( z \right) \right] \frac{e^{- z}}{z} \diff z$. Through extending this result to a stochastic geometry-based settings, the average rate expression in (\ref{InsightRate}) can be expressed as \cite{7277029}
\begin{align}
\mathcal{R} \left(.\right) = \log_{2}(e) \sum_{t^{*} \in \mathcal{T}} \sum_{l^{*} \in \Phi^{(b)}_{t^{*}}} \overline{\varphi}_{t^{*},l^{*},k} \int^{+ \infty}_{0} \int^{+ \infty}_{0} \mathcal{M}_{I_{agg,k}|R} \left( z \right)  \left[ 1 -  \mathcal{M}_{P^{rx}_{t^{*},l^{*},k} | R} \left( z \right) \right]  \frac{e^{-  z \eta }}{z} \mathcal{P}_{\| \hat{Y}_{t^{*},l^{*},k} \|} \left( R \right) \diff z \diff R 
\label{mainER2}
\end{align} 
where $\mathcal{P}_{\| \hat{Y}_{t^{*},l^{*},k} \|}(R)$ is the pdf of the random distance, and $\mathcal{M}_{P^{rx}_{t^{*},l^{*},k} | R}$ and $\mathcal{M}_{I_{agg,k}|R}$ denote the conditional MGFs of the intended signal power and aggregate network interference, respectively. The pdf of transmitter-receiver distance and tier connection probability can be respectively calculated through the analytical expressions \cite{6918448}
\begin{align}
\mathcal{P}_{\| \hat{Y}_{t^{*},l^{*},k} \|}(R) & = \frac{2 \pi R \lambda^{(b)}_{t^{*}} }{\overline{\varphi}_{t^{*},l^{*},k}} \mathbb{E} \left\{ \chi_{t^{*}}^{\frac{2}{ \alpha_{t^{*}}}} \right\} \, e^{- \pi \sum_{t \in \mathcal{T}} \lambda^{(b)}_{t} \mathbb{E} \left\{ \chi_{t}^{\frac{2}{\alpha_{t}}} \right\} \left( \frac{\beta_{t} P^{tx}_{t}}{\beta_{t^{*}} P^{tx}_{t^{*}}} \right)^{\frac{2}{\alpha_{t}}} R^{ \frac{2 \alpha_{t^{*}}}{\alpha_{t}} }}
\label{refdistancee}
\end{align}
and 
\begin{align}
\overline{\varphi}_{t^{*},l^{*},k} & = 2 \pi \lambda^{(b)}_{t^{*}} \mathbb{E} \left\{ \chi_{t^{*}}^{\frac{2}{ \alpha_{t^{*}}}} \right\} \int^{+ \infty}_{0}  r e^{- \pi \sum_{t \in \mathcal{T}} \lambda^{(b)}_{t} \mathbb{E} \left\{ \chi_{t}^{ \frac{2}{\alpha_{t}}} \right\} \left( \frac{\beta_{t} P^{tx}_{t}}{\beta_{t^{*}} P^{tx}_{t^{*}}} \right)^{\frac{2}{\alpha_{t}}} r^{ \frac{2 \alpha_{t^{*}}}{\alpha_{t}} } } \diff r \overset{(a)}{=}
\frac{\lambda^{(b)}_{t^{*}} \mathbb{E} \left\{ \chi_{t^{*}}^{\frac{2}{ \alpha_{t^{*}}}} \right\} }{ \sum_{t \in \mathcal{T}} \lambda^{(b)}_{t} \mathbb{E} \left\{ \chi_{t}^{\frac{2}{\alpha_{t}}} \right\} \left( \frac{\beta_{t} P^{tx}_{t}}{\beta_{t^{*}} P^{tx}_{t^{*}}} \right)^{\frac{2}{\alpha_{t}}}} 
\label{tierconnectt}
\end{align}  
where $(a)$ follows from cases with equivalent path-loss exponent across all different tiers. In addition, the MGF of the intended signal power over Nakagami-m fading channels can be easily computed using
\begin{align}
\mathcal{M}_{P^{rx}_{t^{*},l^{*},k} | R} \left(z \right) = \mathbb{E}_{H_{t^{*},l^{*},k}} \left\{ e^{ -z P^{tx}_{t^{*}} H_{t^{*},l^{*},k} R^{- \alpha_{t^{*}}}} \right\} = \left( 1 + \frac{z P^{tx}_{t^{*}} R^{- \alpha_{t^{*}}}}{m_{t^{*}}} \right)^{-m_{t^{*}}}.
\label{MGF1}
\end{align}
Furthermore, we can derive a closed-form bounded expression for the aggregate network interference MGF - considering the inherent \textit{spatial-correlations} in the activities of \textit{load-proportional} BSs - as in the following lemma. 

\begin{lem}
\textit{The aggregate network interference MGF in spatially-correlated load-proportional heterogeneous DenseNets over Nakagami-m fading interference channels is given by}
\begin{align}
& \tilde{\mathcal{M}}_{I_{agg,k} | R} \left(z \right) = e^{- \displaystyle\pi \displaystyle\sum_{t \in \mathcal{T}} \lambda^{(b)}_{t} \mathbb{E} \left\{ \chi_{t}^{\frac{2}{\alpha_{t}}} \right\} \mathcal{A}_{t}}
\label{MGF2}
\end{align}
where 
\begin{multline}
\mathcal{A}_{t} =  \frac{ \Gamma \left(1- \frac{2}{\alpha_{t}} \right) \Gamma \left(m_{t}+ \frac{2}{\alpha_{t}} \right)}{ \Gamma (m_{t})} \left( \frac{z \overline{\varphi}_{t} P^{tx}_{t}}{m_{t}} \right)^{\frac{2}{\alpha_{t}}} + \left( \frac{\beta_{t} P^{tx}_{t} R^{ \alpha_{t^{*}} }}{\beta_{t^{*}} P^{tx}_{t^{*}}} \right)^{\frac{2}{\alpha_{t}}} \left( m_{t}^{m_{t}} {\left( \frac{z \overline{\varphi}_{t} P^{tx}_{t} \beta_{t^{*}} P^{tx}_{t^{*}} }{\beta_{t} P^{tx}_{t} R^{ \alpha_{t^{*}} }}  +{m_{t}} \right)^{{- m_{t}}}}  - 1 \right) \\  - \frac{ m_{t}^{{m_{t}}+1} }{ \left( z \overline{\varphi}_{t} P^{tx}_{t} \right)^{{m_{t}}} \big( {m_{t}} + \frac{2}{\alpha_{t}} \big)} \left( \frac{\beta_{t} P^{tx}_{t} R^{ \alpha_{t^{*}} }}{\beta_{t^{*}} P^{tx}_{t^{*}}} \right)^{{m_{t}} + \frac{2}{\alpha_{t}} } \, _2F_1 \left({m_{t}}+1,{m_{t}}+ \frac{2}{\alpha_{t}} ;{m_{t}}+ \frac{2}{\alpha_{t}} +1;  \frac{-{m_{t}}\beta_{t} R^{ \alpha_{t^{*}} }}{z \overline{\varphi}_{t} {\beta_{t^{*}} P^{tx}_{t^{*}}}} \right)
\end{multline}
and
\begin{align}
\overline{\varphi}_{t} = 1 - e^{\displaystyle\frac{- \lambda^{(u)} \, \overline{\varphi}_{t,l,k}}{ \lambda^{(b)}_{t} \mathbb{E} \Big\{\chi_{t}^{\frac{2}{\alpha_{t}}} \Big\}}}.
\label{LoadFunction}
\end{align}
In the case of Rayleigh fading interference channels, $\mathcal{A}_{t}$ reduces to 
\begin{multline}
\mathcal{A}_{t} = \Gamma \left(1 - \frac{2}{\alpha_{t}} \right) \Gamma \left(1 + \frac{2}{\alpha_{t}} \right) (z \overline{\varphi}_{t} P^{tx}_{t})^{\frac{2}{\alpha_{t}} } - \frac{ 1 }{1 +\frac{ \beta_{t} R^{ \alpha_{t^{*}}  }}{z \overline{\varphi}_{t} {\beta_{t^{*}} P^{tx}_{t^{*}}}} } \left( \frac{\beta_{t} P^{tx}_{t} R^{ \alpha_{t^{*}}  }}{\beta_{t^{*}} P^{tx}_{t^{*}}} \right)^{\frac{2}{\alpha_{t}}} \\ - \frac{ 1 }{z \overline{\varphi}_{t} P^{tx}_{t} ( 1 + \frac{2}{\alpha_{t}} ) } \left( \frac{\beta_{t} P^{tx}_{t} R^{\alpha_{t^{*}}}}{\beta_{t^{*}} P^{tx}_{t^{*}}} \right)^{1 + \frac{2}{\alpha_{t}}} \, _2F_1\left(2,1 + \frac{2}{\alpha_{t}} ; 2+ \frac{2}{\alpha_{t}} ;\frac{- \beta_{t} R^{\alpha_{t^{*}}  } }{z \overline{\varphi}_{t} {\beta_{t^{*}} P^{tx}_{t^{*}}}} \right).
\end{multline}
For the special case of $\alpha_{t} = 4$, $\forall t \in \mathcal{T}$, the above can be further simplified to
\begin{align}
\mathcal{A}_{t} = \sqrt{z \overline{\varphi}_{t} P^{tx}_{t} } \left( \arctan\left(R^2 \sqrt{ \frac{\beta_{t}}{ z \overline{\varphi}_{t} \beta_{t^{*}} P^{tx}_{t^{*}} }} \right) - \frac{\pi}{2} \right).
\end{align}
\textit{Proof:} See Appendix \ref{BoundedMGFofInt}. 
\end{lem}

Adopting the proposed generalized analytical framework allows for the efficient computation of average rate bound $\tilde{\mathcal{R}}(.)$ in spatially-correlated load-proportional multi-tier cellular networks through double integral operations (compared to the manifold integrals the direct pdf-based approach requires). For homogeneous DenseNet setups, $\tilde{\mathcal{R}}(.)$ can be expressed in a single-integral format when considering interference-limited Rayleigh fading channels, as illustrated in the following lemma. Note that the tier index $t$ is accordingly removed from the system parameters for single-tier cases. 

\begin{lem}
\textit{The average rate bound for interference-limited homogeneous DenseNets over Rayleigh fading channels can be expressed as}
\begin{align}
\tilde{\mathcal{R}}(.) & = \log_{2}(e) \int_0^{\frac{\pi }{2}} \frac{ 2 \alpha^2 (\alpha +2)  \tan (t) \cos ^{2 \alpha}(t)}{\big(\overline{\varphi} \sin ^{\alpha }(t)+\cos ^{\alpha }(t) \big) \left( \alpha  \sin^{\alpha+2}(t) \, _2F_1\left(1,1 + \frac{2}{\alpha };2+\frac{2}{\alpha };-\tan^{\alpha}(t)\right)+\frac{\pi}{2}  (\alpha +2) \csc \left(\frac{2 \pi }{\alpha }\right) \cos^{\alpha + 2}(t) \right) } \diff t. 
\label{SRate0}
\end{align}
\textit{where} $\csc \left(\frac{2 \pi }{\beta }\right) = \frac{\beta}{2 \pi } \Gamma \left(1 - \frac{2}{\beta }\right)  \Gamma \left(1+\frac{2}{\beta }\right)$.
\textit{The above integral can be further simplified for the special case of path-loss exponent being equal to four as }
\begin{align}
\tilde{\mathcal{R}}(.) = \log_{2}(e) \int_0^{+ \infty } \frac{4}{\left( 1 + \overline{\varphi} s^2 \right) \left(2 s - 2 \arctan(s) + \pi \right)} \diff s
\label{SRateLower}
\end{align}
\textit{or alternatively}
\begin{align}
\tilde{\mathcal{R}}(.) = \log_{2}(e) \int_0^{+ \infty } \frac{4}{2 \left( 1 + \overline{\varphi} s^2 \right) \left(s + \arctan(\frac{1}{s}) \right)} \diff s
\label{SRateUpper}
\end{align}
\textit{where}
\begin{align}
\overline{\varphi} = 1 - e^{ \frac{- \lambda^{(u)}}{ \lambda^{(b)} \mathbb{E} \left\{\chi^{\frac{2}{\alpha}} \right\}}}.
\end{align}
\textit{Proof:} See Appendix \ref{SimplifiedRateExpression}.
\end{lem}

It should be elaborated that the different equivalent expressions of $\tilde{\mathcal{R}}(.)$ in (\ref{SRateLower}) and (\ref{SRateUpper}) are provided for the sake of deriving bounded expressions for the optimal deployment solution. This will be elucidated as analysis proceeds.  

\section{Optimal Deployment Solution}
\label{eeSec}

From the mobile operators point of view, the commercial viability of network densification depends on the underlying capital and operational expenditure \cite{6664198}. While the former cost may be covered by taking up a high volume of customers, with the rapid rise in the price of energy, and given that BSs are particularly power-hungry, energy efficiency has become an increasingly crucial factor for the success of DenseNets \cite{7110517}. Generally, there are two main approaches to enhance the energy consumption of cellular networks: (1) improvement in hardware and (2) green system architecture design. Evidently, improving the power consumption of hardware is important. The potential gain, however, will be similar to the business-as-usual case. Hence, the majority of improvement would have to come from green cellular network design. 

In the remaining parts of this paper, we utilize the proposed green analytical framework in order to pinpoint the most energy-efficient deployment solution and hence analyze the achievable gain in energy efficiency by utilizing dynamic sleep modes. The radio planning task under consideration concerns the radio planning task of a service provider for determining the most energy-efficient deployment solution for providing a minimum average rate to its users. 

In order to compute the optimal combination of BS densities towards minimizing total energy expenditure, we formulate the following optimization problem 
\begin{align}
& \underset{\lambda^{(b)}_{t}, t \in \mathcal{T}}{\text{minimize}} \quad \sum_{t \in \mathcal{T}} \mathcal{C}_{t} \lambda^{(b)}_{t} \label{opteq1} \\ & 
\text{subject to:} \quad \mathcal{R} \left(\lambda^{(u)},T,\lambda^{(b)}_{t},P^{tx}_{t},\beta_{t},\eta,\alpha_{t},m_{t},\mu_{t},\sigma_{t} \right) \geq \mathcal{R}_{0} \label{opteq2}
\end{align}  
where $\mathcal{R}_{0}$ is the minimum rate requirement and $\mathcal{C}_{t}$ is the energy expenditure associated with tier-$t$ BSs in active (transmission) mode. Note that $\mathcal{R}$ is a strictly increasing monotonic function in $\lambda^{(b)}_{t}$ (a proof is provided in Appendix \ref{MonotoneFunc}), hence, the optimization problem under consideration has a unique solution. However, $\mathcal{R}$ is a highly complex function which involves for each tier two improper integrals and an infinite series sum. As a result, an exact closed-form solution cannot be obtained. The problem should therefore be tackled numerically. 

Exhaustive search algorithms are well-suited for tackling the problem considering the rate function derivative is not available analytically and its accurate evaluation (e.g., using finite differences) is resource-intensive. In the case of homogeneous DenseNets, the dimension of freedom is reduced to one and the optimization task is equivalent to finding the root of a univariate function. Hence, Brent's algorithm \cite{opac-b1082765} is a reasonable method of choice which at its best (worst) provides super-linear (linear) convergence to the solution. On the other hand, the non-linear constrained multidimensional optimization problem in the case of heterogeneous DenseNets can be tackled using heuristic downhill simplex method \cite{NelderMead65} with penalty function. The algorithm is typically solvable in exponential time \cite{ANAC:ANAC200410015}. The reader is referred to 
\cite{Press:2007:NRE:1403886} for detailed descriptions of the above algorithms operation and code in $C$ language. 

On the other hand, to gain an analytical insight into the effect of different operational settings on the most energy-efficient deployment solution, we focus on the problem of finding the optimal BS density in homogeneous DenseNets. Specifically, we show that it is possible to develop bounded closed-form expressions of the average rate for interference-limited cases with the path-loss exponent being equal to four. 

\begin{lem}
\textit{The lower-bound and upper-bound closed-form expressions of the average rate in interference-limited homogeneous DenseNets over Rayleigh fading channels with path-loss exponent being equal to four are respectively given by}
\begin{align}
\mathcal{\tilde{R}} & = 2 \log_{2} (e) \frac{\left(\pi ^2 \overline{\varphi}^{\frac{3}{2}}+(\pi -2) \pi  \sqrt{\overline{\varphi}} -2 \ln (\pi  \overline{\varphi})+\ln (4)\right) + \frac{2 \pi  \overline{\varphi}+\pi -4}{\sqrt{\pi  (8-\pi )}} \left(2 \arctan \left( \sqrt{\frac{\pi}{\left( 8-\pi \right)}} \right) - \pi \right)}{ \left(\pi ^2 \overline{\varphi}^2+(\pi -4) \pi  \overline{\varphi}+4\right)}
\label{LALB}
\end{align}
and
\begin{align}
\mathcal{\tilde{R}} & = \log_{2} (e) \frac{\pi  \overline{\varphi}^{\frac{3}{2}} + \frac{2}{3 \sqrt{3}} \pi \left( 1 - 2 \overline{\varphi} \right) - \log (\overline{\varphi}) }{\overline{\varphi}^2 - \overline{\varphi} + 1}.
\label{LAUB}
\end{align}
\textit{Proof:} See Appendix \ref{rateApprox}.
\end{lem}

For the above special cases of interest, we derive closed-form solutions for the minimum number of BSs per unit area ${\lambda^{(b)}}^{*}$ needed to satisfy the service requirement based on the numerical roots of high-order polynomial functions. The results, capturing the worst- and best-case scenario of the deployment solution, are respectively presented in the following lemma. 

\begin{lem}
\textit{The bounded solutions of the optimal BS density in interference-limited homogeneous DenseNets over Rayleigh fading channels with path-loss exponent being equal to four are expressed as}
\begin{align}
{\lambda^{(b)}}^{*} & \leq  \lambda^{(u)} \mathbb{E} \Big\{ \chi^{\frac{2}{\alpha}} \Big\} \biggr/ -\ln \Biggr( 1 - \text{positive real root of } \, \biggr\{ \sqrt{\pi  (8-\pi )} \big(4 \left(-x^2-\pi  x+1\right) +2 \pi ^2 x \left(x^2+1\right) \nonumber \\ & - \ln(2) \mathcal{R}_{0} \left( \pi x^2 \left( \pi \left(x^2+1\right) -4 \right) + 4\right) +\ln (16)-4 \ln (\pi ) \big) - 4 \left(2 \pi  x^2+\pi -4\right) \left(\pi -2 \arctan \left( \sqrt{\frac{\pi}{8-\pi }} \right) \right) \biggr\}^2 \Biggr)
\label{optWC}
\end{align}
and
\begin{align}
{\lambda^{(b)}}^{*} & \geq \lambda^{(u)} \mathbb{E} \Big\{ \chi^{\frac{2}{\alpha}} \Big\} \biggr/ - \ln \Biggr(1-\text{positive real root of } \, \biggr\{ 9 \biggr( \pi  \biggr( x^3- \biggr( \frac{4}{3 \sqrt{3}}+\frac{1}{\pi } \biggr) x^2 + \frac{2}{3 \sqrt{3}} \biggr) \nonumber \\ & - \ln(2) \mathcal{R}_{0} \left( x^4-x^2+1 \right) \biggr) \biggr\}^2 \Biggr)
\label{optBC}
\end{align}
\textit{Proof:} See Appendix \ref{OptSolBett}.
\end{lem}

It is important to note from the best and worst BS deployment density expressions, respectively derived in (\ref{optWC}) and (\ref{optBC}), that the optimal network density ${\lambda^{(b)}}^{*}$ is directly proportional to the network load $\lambda^{(u)}$. The tightness of the developed bounded expressions is analyzed and compared to theoretical (using exhaustive search algorithms) and Monte-Carlo simulation results in the next section. Note that the analytical tractability of the techniques used to derive these bounds quickly diminishes considering multi-tier deployments. For heterogeneous DenseNets with equivalent operational parameters across different tiers, i.e., equal energy cost, biasing weight, and transmission power, however, we respectively arrive at similar bounded optimal network density expressions with ${\lambda^{(b)}}^{*} \rightarrow \sum_{t \in \mathcal{T}}{\lambda^{(b)}_{t}}^{*}$.

\section{Power Savings with Sleep Modes}
\label{psSec}

It is well-known that due to the recent advances in hardware technology, it has been made possible for wireless transceivers to consume varying power levels under different operational modes \cite{6619582}. These include BS sleep, idle, transmit, and receive modes which can be accordingly adjusted based on the daily fluctuations in the traffic volume for the purpose of preserving energy. 

Defining a quantitative BS power model is however challenging given that one needs to take into consideration the particular components configurations. The following linear power model is however shown to be a reasonable approximation \cite{6056691}
\begin{align}
P_{t} = 
\begin{cases}
\mathcal{C}_{t} = \Delta^{(p)}_{t} P^{tx}_{t} + P^{(e)}_{t} & \text{if BS is in transmit mode} \\
P^{(s)}_{t} & \text{if BS is in sleep mode} 
\end{cases}
\end{align}
where for tier-$t$ BSs, $\Delta^{(p)}_{t}$ is the reciprocal of the power amplifier drain efficiency, $P^{(e)}_{t}$ is the circuit power, and $P^{(s)}_{t}$ is the power in sleep mode. This power model accounts for the different specifications and architectures of long-term-evolution (LTE) BSs including macro, micro, and pico types. The sleep mode power consumption (when there is nothing to transmit) is also included in this model to reflect upon a promising energy savings mechanism associated with future BSs. A set of power values for a default operating scenario can be found in \cite[TABLE 2]{6056691}.


In order to estimate the optimal energy savings from load-proportional network behavior in a given cellular environment, we utilize the established theoretical framework to identify the number and type of BSs that maintain the rate requirement for the users' density over different times of the day. The power savings in Watts using dynamic sleep modes at a given time of the day can then be computed by utilizing the linear hardware model as follows
\begin{align}
\mathcal{S} = \sum_{t \in \mathcal{T}} \left( {\lambda_{t,f}^{(b)}}^{*} - {\lambda_{t,p}^{(b)}}^{*} \right) \left( \Delta^{(p)}_{t} P^{tx}_{t} + P^{(e)}_{t} - P^{(s)}_{t} \right) 
\end{align}
where ${\lambda_{t,f}^{(b)}}^{*}$ and ${\lambda_{t,p}^{(b)}}^{*}$, $t \in \mathcal{T}$, are the optimal tier-$t$ network densities under full and partial (depending on the hour) network loads, respectively. It is important to note that although this framework cannot determine an optimal topology for a given area, it can provide valuable information concerning how many BSs are required to meet the traffic demand, and in turn, how many BSs can be switched off according to the temporal variations in the traffic volume. 


\section{Performance Evaluation}
\label{secRES}

The aim of this section is to evaluate the average rate, optimal deployment density, and power savings of DenseNets, considering different combinations of large-cell macro and small-cell micro and pico BSs. We aim to quantify the impact of network-wide decisions which helps unveil important design pointers for optimal network management. In regards to the power model, we use the practical hardware values captured in \cite{6056691}. To analyze the accuracy of the established theoretical model, we perform load-dependent Monte-Carlo simulations (see Appendix \ref{Monte-Carlotrials}).
\subsection{Framework Validation and Impact of System Parameters}

\begin{figure}[!t]
\centering
\includegraphics{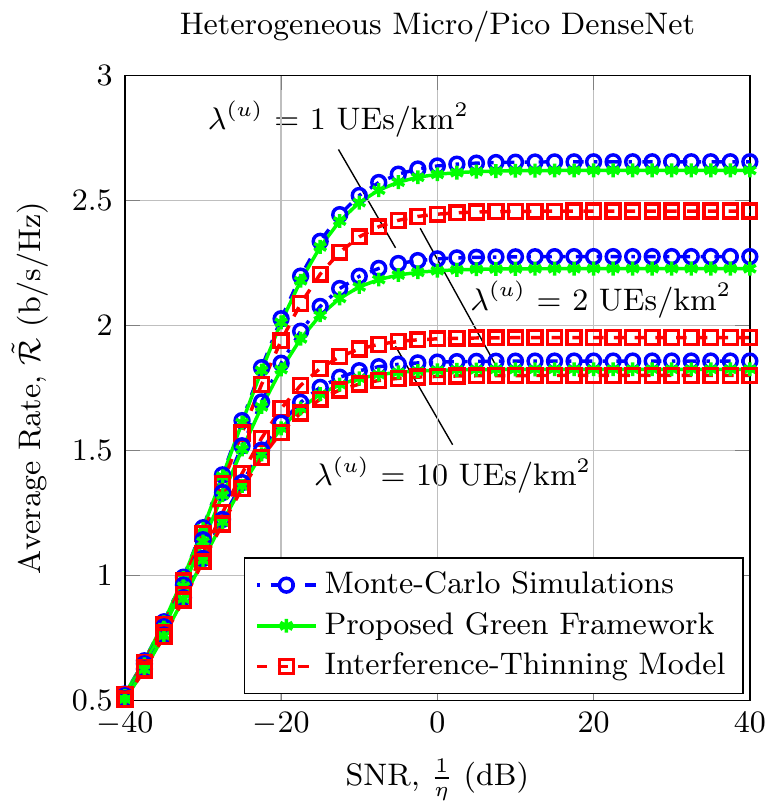}
\caption{Average rate under different network load and noise power values, $\lambda^{(b)}_{m} = \lambda^{(b)}_{p}  =  2$ BSs/km$^{\text{2}}$, $P^{tx}_{m} = 6.3$ W, $P^{tx}_{p} = 0.13$ W, $\beta_{m} = \beta_{p} = 0$ dB, $m_{m}  = m_{p} = 1$, $\alpha_{m} = \alpha_{p}  =  4$.}
\label{HetNetRate}
\end{figure}

\begin{figure}[!t]
\centering
\includegraphics{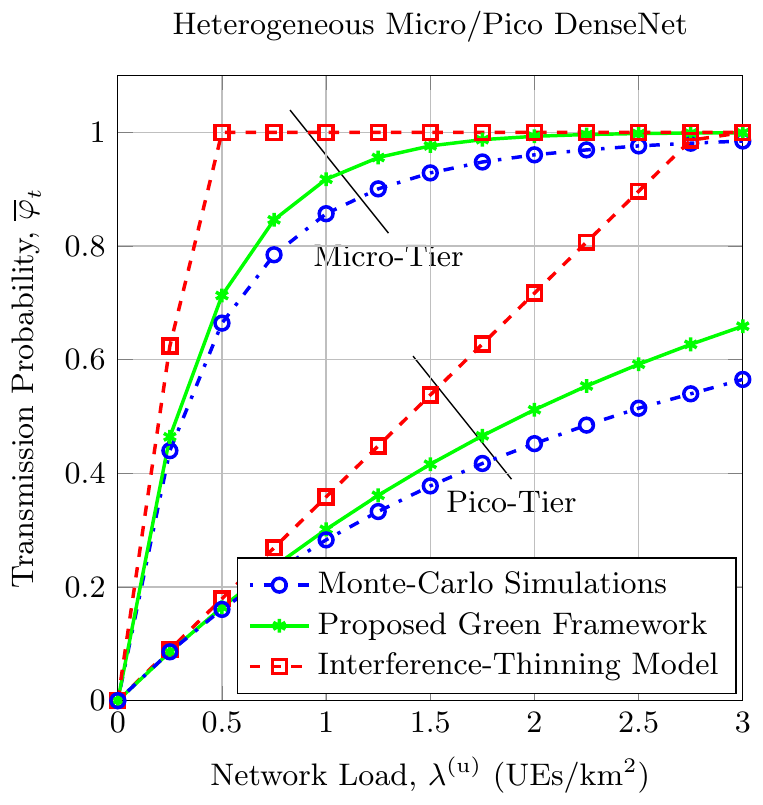}
\caption{Transmission probability as a function of network load, $\lambda^{(b)}_{m}  =  0.3$ BSs/km$^{\text{2}}$, $\lambda^{(b)}_{p}  =  0.7$ BSs/km$^{\text{2}}$, $P^{tx}_{m} = 6.3$ W, $P^{tx}_{p} = 0.13$ W, $\beta_{m} = \beta_{p} = 0$ dB, SNR $=  20$ dB, $m_{m} = m_{p}  =  1$, $\alpha_{m} = \alpha_{p} = 4$.}
\label{CellNetTransmissionProbabiliy}
\end{figure}

The performances of a mixed micro/pico system under different SNR and load levels using the interference-thinning model, proposed green framework, and Monte-Carlo simulations are shown in Fig. \ref{HetNetRate}. A key point to observe is that the former approach, while being an improvement over the long-standing fully-loaded model, produces pessimistic performance values, particularly under light and moderate network loads. Furthermore, our proposed analytical model correctly provides a tight lower-bound fit of the actual performance curve. The gap between different evaluation tools is negligible under both heavy traffic, due to the full-loaded interference field, and low SNRs, due to noise power dominance over interference. To further illustrate the shortcomings of the uncorrelated interferers assumption, we plot the transmission probability of the micro and pico tiers in a heterogeneous DenseNet as a function of network load in Fig. \ref{CellNetTransmissionProbabiliy}. 

\begin{figure}[!t]
\includegraphics{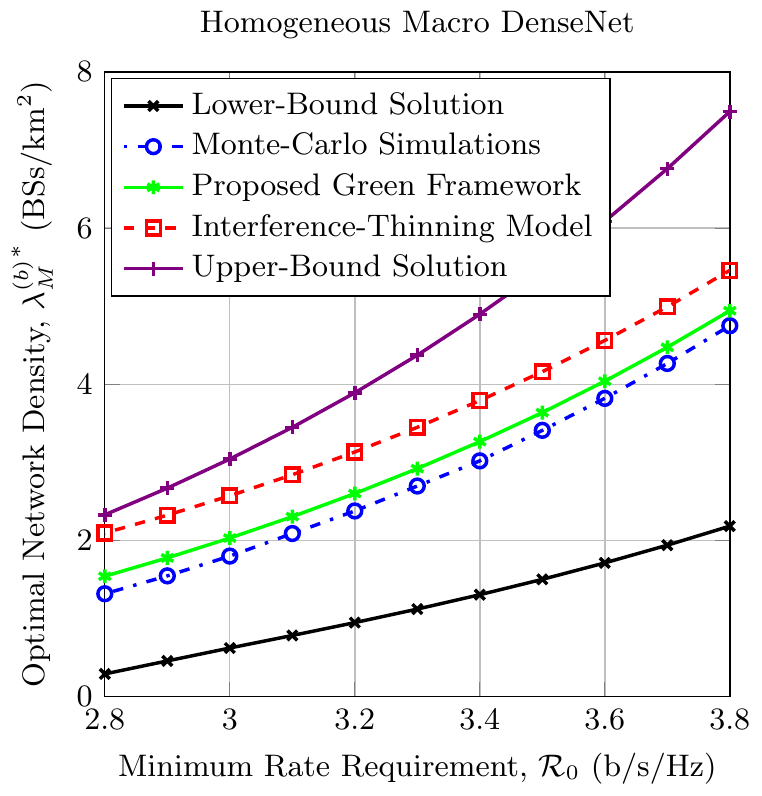}
\centering
\caption{Optimal BS density under different minimum rate requirements, $\lambda^{(u)}  =  1$ UEs/km$^{\text{2}}$, $P^{tx}_{M} = 20$ W, $\eta = 0$ W, $m_{M}  =  1$, $\alpha_{M}  =  4$.}
\label{HomNetwithDensities}
\end{figure}

The implication of the above trends on network cost is depicted in Fig. \ref{HomNetwithDensities}, where we calculate the optimal network density of an interference-limited macro-only system as a function of minimum rate demand using the interference-thinning model, the proposed green framework, Monte-Carlo trials, and the bounded approximations derived in eqs. (\ref{optWC}) and (\ref{optBC}). The applicability of our proposed bounded framework in capturing realistic scenarios is further confirmed as it provides a tight fit to the Monte-Carlo trials. The interference-thinning model, on the other hand, requires a much larger BS density to meet a particular rate requirement. E.g., from Fig. \ref{HomNetwithDensities}, to satisfy $\mathcal{R}_{0} =$ 2.4 b/s/Hz, the interference-thinning model requires 10.7\%, 82.0\%, and 159.0\% larger network density over the closed-form upper-bound solution, proposed green framework (with exhaustive search algorithms), and Monte-Carlo trials, respectively. Henceforth, the analysis is carried out using the proposed framework as we have extensively shown its advantages over the state-of-the-art models.

\begin{figure}[!t]
\centering
\includegraphics{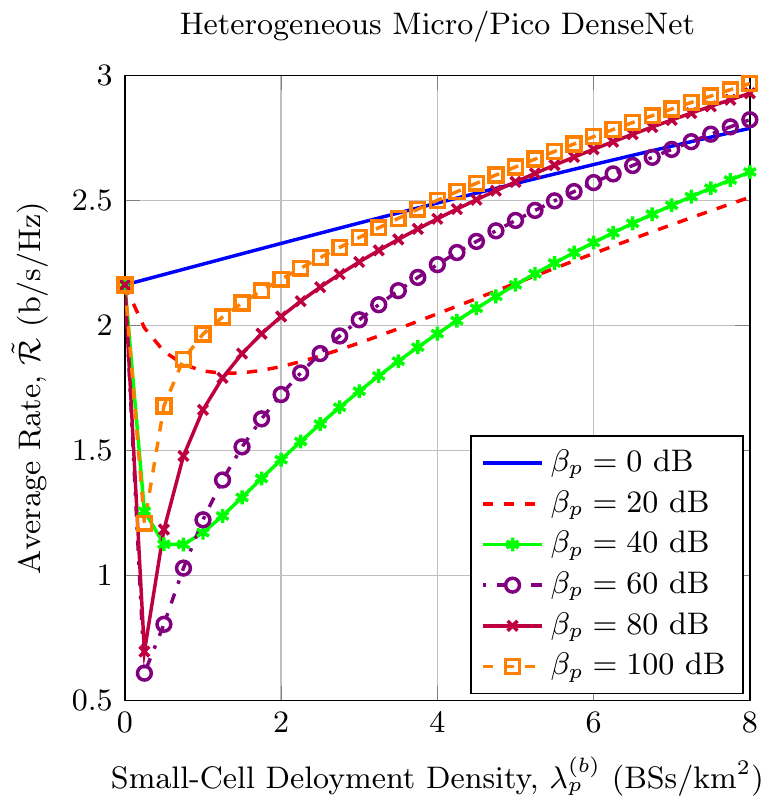}
\caption{Impact of network densification and biasing on average rate performance, $\lambda^{(u)} = 4$ UEs/km$^{\text{2}}$, $\lambda^{(b)}_{m}  =  1$ BSs/km$^{\text{2}}$, $P^{tx}_{m} = 6.3$ W, $P^{tx}_{p} = 0.13$ W, $\beta_{m} = 0$ dB, SNR$ = 25$ dB, $m_{m} = m_{p}  =  1$, $\alpha_{m} = \alpha_{p} = 4$.}
\label{BiasnDense}
\end{figure}

\begin{figure}[!t]
\centering
\includegraphics{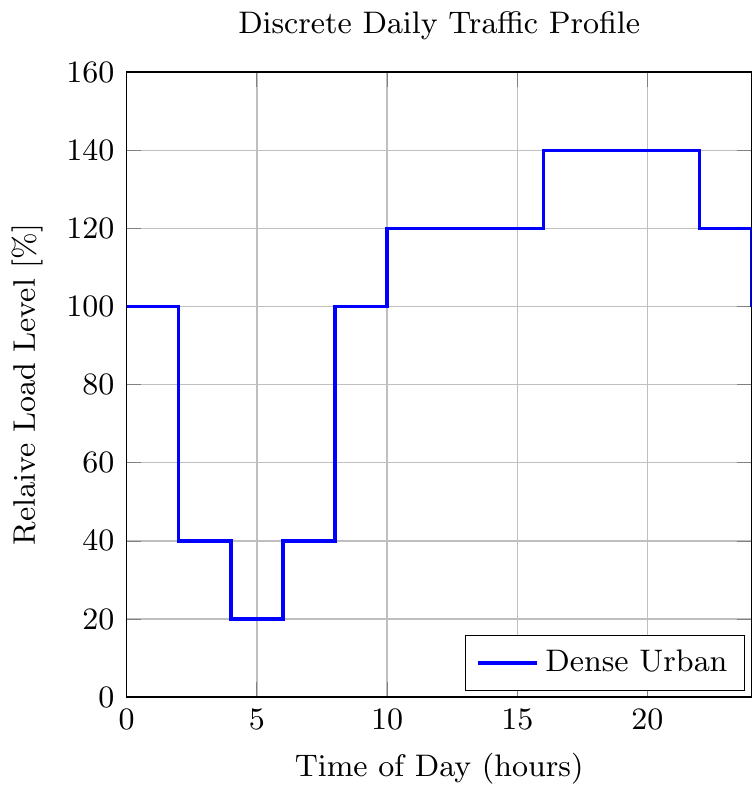}
\caption{Relative load level at different times of the day in a dense urban environment.}
\label{DailyTrafficProfile}
\end{figure}

\begin{figure}[!t]
\centering
\includegraphics{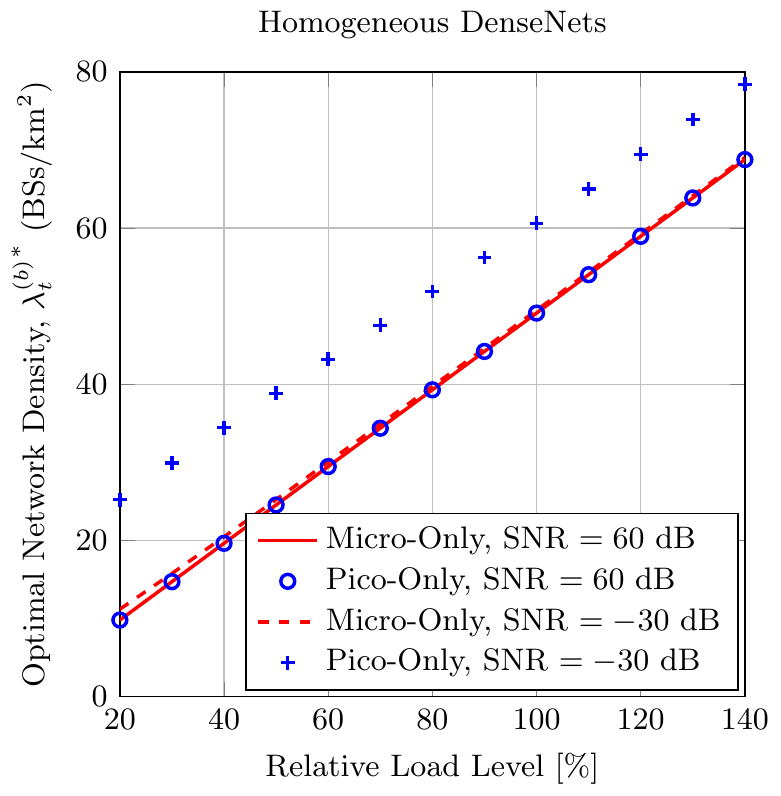}
\caption{Optimal BS densities of different homogeneous DenseNets under different relative load levels, $P^{tx}_{m} = 6.3$ W, $P^{tx}_{p} = 0.13$ W, $m_{m} = m_{p}  =  2$, $\mu_{m} = \mu_{p} = 0$ dB, $\sigma^2_{m} = \sigma^2_{p} = 6$ dB, $\alpha_{m} = \alpha_{p} = 4$, $\mathcal{R}_{0} = 2.4$ b/s/Hz.}
\label{DailyNetworkDensity}
\end{figure}

Before proceeding to the optimal deployment solution analysis, we depict the impact of network densification using small-cells with different biasing values on average rate performance in Fig. \ref{BiasnDense}. Firstly, we observe that performance improves nearly linearly by adding unbiased pico BSs, further confirming the promising potential of small-cell solution in offloading traffic from large-cells in congested areas. However, deploying small-cells with low artificial-bias deteriorates performance. The reason lies on the added intra-tier interference experienced by pico BSs without a significant reduction in the inter-tier interference from the micro BSs.
We can thus infer that the performance gain from artificial expansion of small-cells range is negative or otherwise negligible under relative moderate and heavy traffic loads.  

\begin{table}
\captionsetup{font=small}
\centering
{\small
\begin{tabular}{ccc}
\toprule
Relative Load (\%) & ${\lambda^{(b)}_{m}}^{*}$ (Micro BSs/km$^{\text{2}}$)   & ${\lambda^{(b)}_{p}}^{*}$ (Pico BSs/km$^{\text{2}}$)\\
\midrule
20 & 0.292901 & 7.32528 \tabularnewline
30 & 0.446714 & 11.1706 \tabularnewline
40 & 0.585807 & 14.6506 \tabularnewline
50 & 0.732258 & 18.3132 \tabularnewline
60 & 0.878703 & 21.9760 \tabularnewline
70 & 1.02517 & 25.6385 \tabularnewline
80 & 1.17162 & 29.3012 \tabularnewline
90 & 1.31799 & 32.9648 \tabularnewline
100 & 1.46450 & 36.6269 \tabularnewline
110 & 1.61097 & 40.2894 \tabularnewline
120 & 1.75743 & 43.9520 \tabularnewline
130 & 1.90388 & 47.6147 \tabularnewline
140 & 2.05033 & 51.2774 \tabularnewline
\end{tabular}}
\caption{Optimal densities of BSs in a mixed micro/pico deployment under relative load values, $P^{tx}_{m} = 6.3$ W, $P^{tx}_{p} = 0.13$ W, $P^{(e)}_{m} = 53$ W, $P^{(e)}_{p} = 6.8$ W, $\Delta^{(p)}_{m} = 3.1$, $\Delta^{(p)}_{p} = 4$, SNR $= 60$ dB, $m_{m} = m_{p}  =  2$, $\mu_{m} = \mu_{p} = 0$ dB, $\sigma^2_{m} = \sigma^2_{p} = 6$ dB, $\beta_{m} = \beta_{p} = 0$ dB, $\alpha_{m} = \alpha_{p} = 4$, $\mathcal{R}_{0} = 2.4$ b/s/Hz.}
\label{DenMultiTier}
\end{table}

\begin{figure}[!t]
\centering
\includegraphics{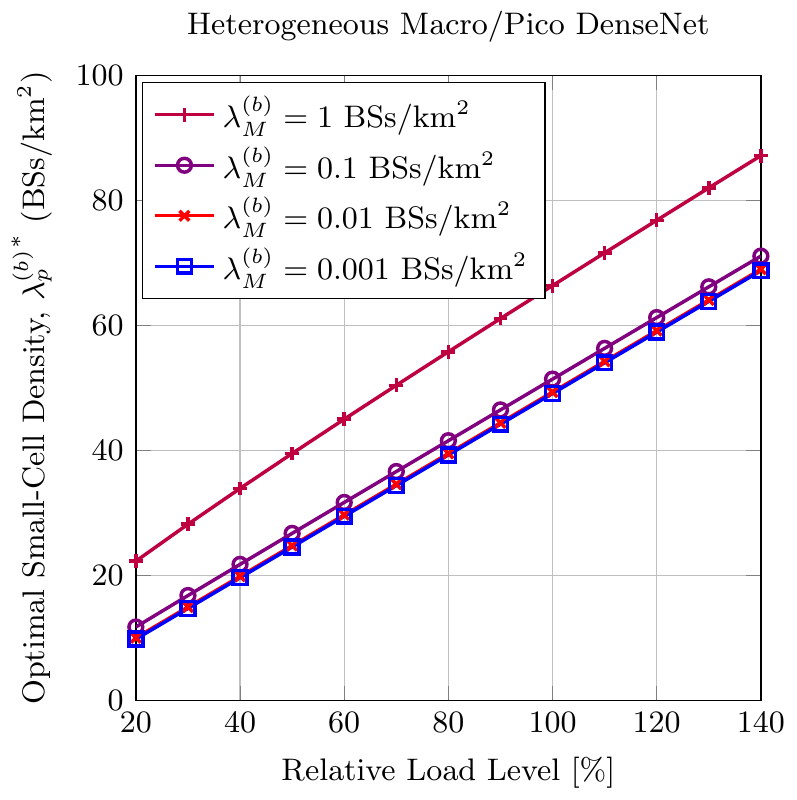}
\caption{Optimal pico BSs density in heterogeneous DenseNets overlaid with different deployment densities of legacy macro-cells under relative load levels, $P^{tx}_{M} = 20$ W, $P^{tx}_{p} = 0.13$ W, $\beta_{M} = \beta_{p} = 0$ dB, $m_{M} = m_{p}  =  2$, $\mu_{M} = \mu_{p} = 0$ dB, $\sigma^2_{M} = \sigma^2_{p} = 6$ dB, $\alpha_{M} = \alpha_{p} = 4$, $\mathcal{R}_{0} = 2.4$ b/s/Hz.}
\label{PracticallyOptimalSolution}
\end{figure}

\begin{figure}[t]
\centering
\includegraphics{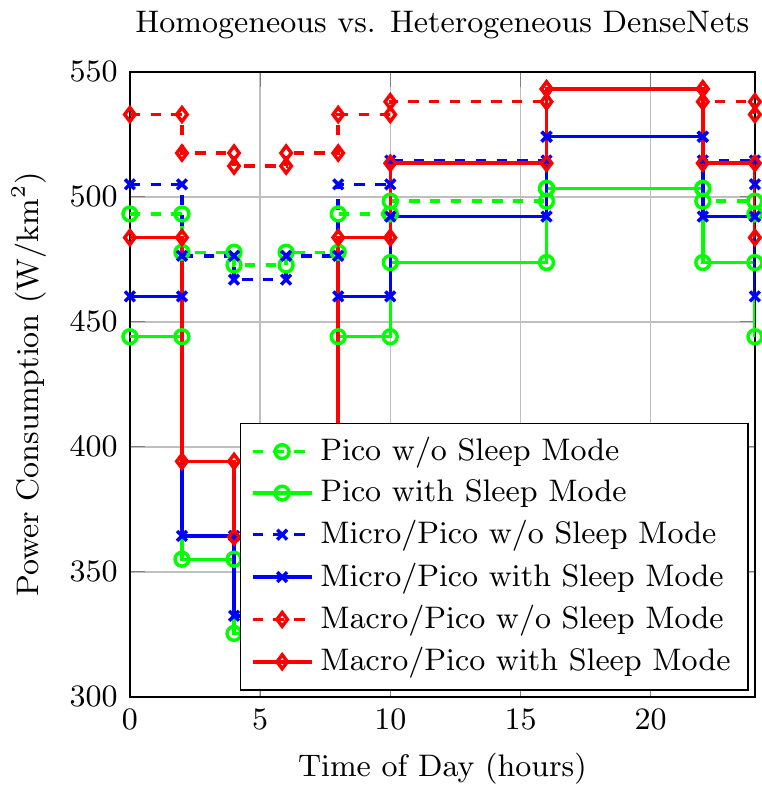}
\caption{Total network power consumption at different times of day, $P^{tx}_{M} = 20$ W, $P^{tx}_{m} = 6.3$ W, $P^{tx}_{p} = 0.13$ W, $P^{(e)}_{M} = 118.7$ W, $P^{(e)}_{m} = 53$ W, $P^{(e)}_{p} = 6.8$ W, $\Delta^{(p)}_{M} = 5.32$, $\Delta^{(p)}_{m} = 3.1$, $\Delta^{(p)}_{p} = 4$, $P^{(s)}_{M} = 93$ W, $P^{(s)}_{m} = 39$ W, $P^{(s)}_{p} = 4.3$ W, $\beta_{M} = \beta_{m} = \beta_{p} = 0$ dB, SNR $= 60$ dB, $m_{M} = m_{m} = m_{p}  =  2$, $\mu_{M} = \mu_{m} = \mu_{p} = 0$ dB, $\sigma^2_{M} = \sigma^2_{m} = \sigma^2_{p} = 6$ dB, $\alpha_{M} = \alpha_{m} = \alpha_{p} = 4$, $\mathcal{R}_{0} = 2.4$ b/s/Hz.}
\label{PowerCompare}
\end{figure}

\begin{figure}[t]
\centering
\includegraphics{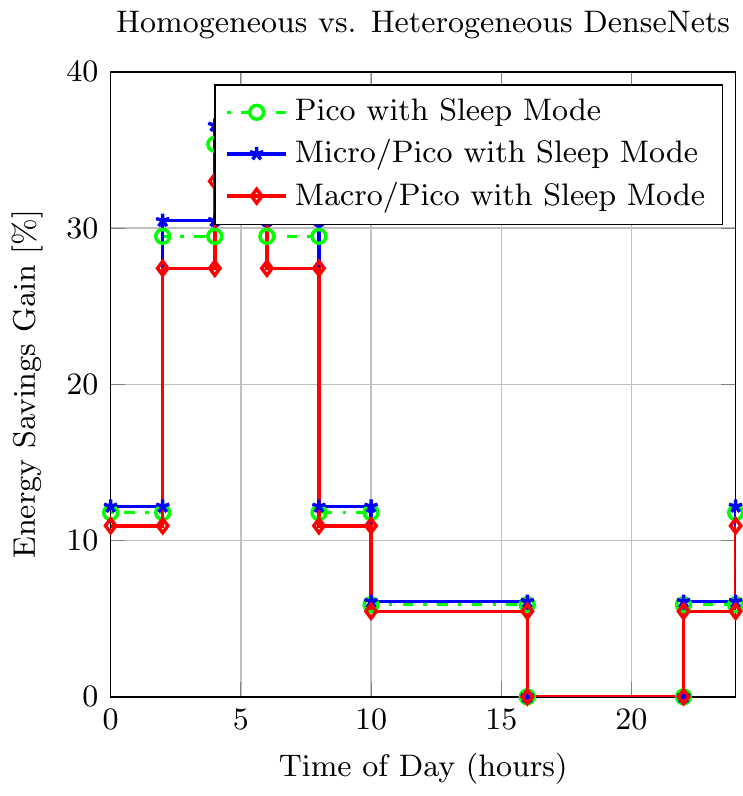}
\caption{Relative energy savings gain from dynamic BS switching at different times of day, $P^{tx}_{M} = 20$ W, $P^{tx}_{m} = 6.3$ W, $P^{tx}_{p} = 0.13$ W, $P^{(e)}_{M} = 118.7$ W, $P^{(e)}_{m} = 53$ W, $P^{(e)}_{p} = 6.8$ W, $\Delta^{(p)}_{M} = 5.32$, $\Delta^{(p)}_{m} = 3.1$, $\Delta^{(p)}_{p} = 4$, $P^{(s)}_{M} = 93$ W, $P^{(s)}_{m} = 39$ W, $P^{(s)}_{p} = 4.3$ W, $\beta_{M} = \beta_{m} = \beta_{p} = 0$ dB, SNR $= 60$ dB, $m_{M} = m_{m} = m_{p}  =  2$, $\mu_{M} = \mu_{m} = \mu_{p} = 0$ dB, $\sigma^2_{M} = \sigma^2_{m} = \sigma^2_{p} = 6$ dB, $\alpha_{M} = \alpha_{m} = \alpha_{p} = 4$, $\mathcal{R}_{0} = 2.4$ b/s/Hz.}
\label{EnergyCompare}
\end{figure}

\subsection{Optimal Deployment Solution}

In this part, we investigate the optimal deployment solution under traffic and rate requirements anticipated for future dense urban environments. According to the parameters developed for mature markets within GreenTouch, see \cite{6533307} and \cite{6629714}, the anticipated peak traffic volume in the busy hour of a dense urban region in 2020 is 702 Mbits/sec/km$^{\text{2}}$. Further, a discrete daily traffic profile ranging from 20\% to 140\% of the average load in dense urban environments is utilized \cite{6629714}, see Fig. \ref{DailyTrafficProfile}. 
Using the above parameter values, the active UEs density at 100\% load level is calculated to be 84.87 UEs/km$^{\text{2}}$. The recent Ofcom Market Report states that the average rate requirement for users on fourth-generation (4G) services in 2020 is expected to reach 2.4 b/s/Hz.  
Moreover, typical propagation values in dense urban environments are selected with Nakagami-m fading with $m =$ 2 (i.e., two-antenna transmit diversity Rayleigh), Log-Normal shadowing with $\mu_{t} =$ 0 and variance $\sigma^{2}_{t} = $ 6, and path-loss exponent $\alpha_{t} =$ 4.     

The optimal BS densities, which meet users demand under different relative load levels in high and low SNR operating regions, of two different single-tier micro and pico DenseNets are depicted in Fig. \ref{DailyNetworkDensity}. The optimal network density is shown to vary significantly in different hours, e.g., for the noise-dominant case of the micro-only DenseNet, the optimal BS density at 20\% load corresponding to 4-6 am (morning time) is 90.14\% less than the value at 140\% load experienced between 4-10 pm (night time). Furthermore, because of the negligible impact of transmit power in the interference-dominant cases, the optimal deployment density in the two different micro-only and pico-only systems are almost the same. This trend highlights the major disadvantage of deploying energy-hungry large-cells in dense interference-dominant scenarios of the future. On the other hand, in noise-dominant regions, a greater number of pico BSs is required to meet the service requirements, e.g., at 20\% load, ${\lambda^{(b)}}^{*}$ of the pico-only system is 2.94 times greater than the optimal BS density of the single-tier micro deployment. 

We now turn to the more challenging problem of optimal deployment solution in multi-tier DenseNets. Considering all combinations of macro, micro, and pico BSs, we employ an exhaustive search algorithm to compute the most energy-efficient deployment solution, e.g., the ratio of the energy cost in transmission mode of a micro BS over a pico BS is $\frac{\mathcal{C}_{m}}{\mathcal{C}_{p}} = $ 9.8085 and a macro BS over a pico BS is $\frac{\mathcal{C}_{M}}{\mathcal{C}_{p}} = $ 30.7514. Our findings interestingly reveal that a small-cell deployment with only pico BSs with the values previously provided in Fig. \ref{DailyNetworkDensity} is the optimal solution for minimizing total energy expenditure. For comparison purposes, we identify the second most energy-efficient deployment solution as a heterogeneous sparse micro and dense pico network with the values provided TABLE \ref{DenMultiTier}. E.g., under a relative load of 100\%, with approximately 1.464 micro BSs/km$^{\text{2}}$ BSs, the heterogeneous micro/pico DenseNet requires 1.34 times fewer pico-cells for satisfying $\mathcal{R}_{0} =$ 2.4 b/s/Hz over the optimal homogeneous pico DenseNet. Note that by further reducing the noise power down to zero watts, we record no significant changes to the optimal type and number of BSs obtained.

Note that the practical feasibility of the above solutions is justified given the already in place legacy cellular networks are utilized for satisfying the global coverage constraint. However, a ``practically optimal" solution could arguably accommodate the existence of legacy macro-cells in both data and control planes. In Fig. \ref{PracticallyOptimalSolution}, we depict the optimal pico BSs density needed to satisfy the rate requirement in mixture legacy cellular networks overlaid with different amounts of macro-cells. We observe that the use of high-power macro-cells in the data plane for dense urban environments has a detrimental impact considering a higher amount of small-cells is needed to meet the average rate constraint. For example, under a relative maximum load of 140\%, the optimal deployment density of small-cells is around 3.43\% lower in a homogeneous pico system over a mix macro/pico DenseNet with $\lambda^{(b)}_{M} =$ 0.1 BSs/km$^{\text{2}}$. It can therefore be inferred that in such environments the use of existing legacy cellular networks must be directed solely towards providing global coverage. 

\subsection{Power Savings via Sleep Modes}

Finally, we analyze and compare the total power consumption and energy savings gain of different deployment scenarios in the dense urban environment under consideration. Fig. \ref{PowerCompare} depicts the power consumed per unit area at different times of the day considering different networks equipped with and without BS sleep modes. The results confirm the optimality of the homogeneous pico DenseNet in terms of total energy efficiency, e.g., at peak traffic level, the mixed micro/pico deployment consumes more than 4.11\% power than the optimal solution. Both of these solutions, however, are considerably more energy-efficient compared to any other combinations of BS densities such as the mix macro/pico system. E.g., the optimal pico DenseNet is capable of realizing peak power savings of near 15 kW/km$^{\text{2}}$ over a stand-alone macro-cell deployment; additional 20 W/km$^{\text{2}}$ and 40 W/km$^{\text{2}}$ improvements compared to the heterogeneous micro/pico and macro/pico DenseNets, respectively. It can also be seen from Fig. \ref{PowerCompare} that due to the large difference in the power usage of idle and sleep states, a significant reduction in total energy consumption can be achieved by activating BSs on-demand. Specifically, as shown in Fig. \ref{EnergyCompare}, by powering down BSs, relative to operating under full-load, peak energy efficiency gains of 35.36\%, 36.57\%, and 33.0\% at night time, and average daily gains of 11.78\%, 12.19\%, and 10.97\% for the pico-only, mixed micro/pico, and mixed macro/pico systems are respectively recorded.   


\section{Conclusions}
\label{secCON}

We have provided a comprehensive theoretical framework for performance evaluation and optimization of dense cellular networks. By incorporating the notion of load-proportionality in a flexible separated data/control plane cellular network architecture,
we identify the most energy-efficient deployment solution for for satisfying a minimum rate requirement under a given traffic level. The validity of the proposed green framework and its advantages over state-of-the-art fully-loaded and interference-thinning models in terms of pinpointing the optimal deployment density required for meeting users' demands was confirmed through extensive Monte-Carlo trials. Under a dense urban environment in the year 2020, the optimal deployment solution for the capacity plane was found to be a populated homogeneous pico network capable of realizing power savings of upto near 15 kW/km$^{\text{2}}$ compared to a traditional stand-alone macro cellular network.  

\bibliographystyle{IEEEtran}
\bibliography{IEEEabrv,myref}

\begin{thebibliography}{10}
\providecommand{\url}[1]{#1}
\csname url@samestyle\endcsname
\providecommand{\newblock}{\relax}
\providecommand{\bibinfo}[2]{#2}
\providecommand{\BIBentrySTDinterwordspacing}{\spaceskip=0pt\relax}
\providecommand{\BIBentryALTinterwordstretchfactor}{4}
\providecommand{\BIBentryALTinterwordspacing}{\spaceskip=\fontdimen2\font plus
\BIBentryALTinterwordstretchfactor\fontdimen3\font minus
  \fontdimen4\font\relax}
\providecommand{\BIBforeignlanguage}[2]{{%
\expandafter\ifx\csname l@#1\endcsname\relax
\typeout{** WARNING: IEEEtran.bst: No hyphenation pattern has been}%
\typeout{** loaded for the language `#1'. Using the pattern for}%
\typeout{** the default language instead.}%
\else
\language=\csname l@#1\endcsname
\fi
#2}}
\providecommand{\BIBdecl}{\relax}
\BIBdecl

\bibitem{7504384}
A.~Shojaeifard, K.~A. Hamdi, E.~Alsusa, D.~K.~C. So, and J.~Tang, ``Optimal
  deployment of dense cellular networks,'' in \emph{2016 IEEE 83rd Veh.
  Technol. Conf. (VTC Spring)}, May 2016, pp. 1--5.

\bibitem{6736747}
N.~Bhushan, J.~Li, D.~Malladi, R.~Gilmore, D.~Brenner, A.~Damnjanovic,
  R.~Sukhavasi, C.~Patel, and S.~Geirhofer, ``Network densification: the
  dominant theme for wireless evolution into {5G},'' \emph{IEEE Commun. Mag.},
  vol.~52, no.~2, pp. 82--89, Feb. 2014.

\bibitem{6477048}
H.~ElSawy and E.~Hossain, ``Two-tier {HetNets} with cognitive femtocells:
  Downlink performance modeling and analysis in a multichannel environment,''
  \emph{IEEE Trans. Mobile Comput.}, vol.~13, no.~3, pp. 649--663, Mar. 2014.

\bibitem{7118688}
J.~Tang, D.~K.~C. So, E.~Alsusa, K.~A. Hamdi, and A.~Shojaeifard, ``Energy
  efficiency optimization with interference alignment in multi-cell {MIMO}
  interfering broadcast channels,'' \emph{IEEE Trans. Commun.}, vol.~63, no.~7,
  pp. 2486--2499, July 2015.

\bibitem{5978416}
A.~Fehske, G.~Fettweis, J.~Malmodin, and G.~Biczok, ``The global footprint of
  mobile communications: The ecological and economic perspective,'' \emph{IEEE
  Commun. Mag.}, vol.~49, no.~8, pp. 55--62, Aug. 2011.

\bibitem{7177127}
J.~Tang, D.~K.~C. So, E.~Alsusa, K.~A. Hamdi, and A.~Shojaeifard, ``On the
  energy efficiency-spectral efficiency tradeoff in {MIMO}-{OFDMA} broadcast
  channels,'' \emph{IEEE Trans. Veh. Technol.}, vol.~65, no.~7, pp. 5185--5199,
  July 2016.

\bibitem{101}
GreenTouch initiative. www.greentouch.org.

\bibitem{6171996}
H.~Dhillon, R.~Ganti, F.~Baccelli, and J.~Andrews, ``Modeling and analysis of
  {K}-tier downlink heterogeneous cellular networks,'' \emph{IEEE J. Sel. Areas
  Commun.}, vol.~30, no.~3, pp. 550--560, Apr. 2012.

\bibitem{6126035}
R.~Ganti, F.~Baccelli, and J.~Andrews, ``Series expansion for interference in
  wireless networks,'' \emph{IEEE Trans. Inform. Theory}, vol.~58, no.~4, pp.
  2194--2205, Apr. 2012.

\bibitem{6365639}
G.~Alfano, M.~Garetto, and E.~Leonardi, ``New directions into the stochastic
  geometry analysis of dense {CSMA} networks,'' \emph{IEEE Trans. Mobile
  Comput.}, vol.~13, no.~2, pp. 324--336, Feb. 2014.

\bibitem{6516171}
M.~Di~Renzo, A.~Guidotti, and G.~Corazza, ``Average rate of downlink
  heterogeneous cellular networks over generalized fading channels: A
  stochastic geometry approach,'' \emph{IEEE Trans. Commun.}, vol.~61, no.~7,
  pp. 3050--3071, Jul. 2013.

\bibitem{7277029}
A.~Shojaeifard, K.~Hamdi, E.~Alsusa, D.~So, and J.~Tang, ``Exact {SINR}
  statistics in the presence of heterogeneous interferers,'' \emph{IEEE Trans.
  Inform. Theory}, vol.~61, no.~12, pp. 6759--6773, Dec. 2015.

\bibitem{6524460}
H.~Elsawy, E.~Hossain, and M.~Haenggi, ``Stochastic geometry for modeling,
  analysis, and design of multi-tier and cognitive cellular wireless networks:
  A survey,'' \emph{IEEE Commun. Surveys Tuts.}, vol.~15, no.~3, pp. 996--1019,
  Jun. 2013.

\bibitem{5223626}
M.~Di~Renzo, F.~Graziosi, and F.~Santucci, ``Channel capacity over generalized
  fading channels: A novel {MGF}-based approach for performance analysis and
  design of wireless communication systems,'' \emph{IEEE Trans. Veh. Technol.},
  vol.~59, no.~1, pp. 127--149, Jan. 2010.

\bibitem{5407601}
K.~Hamdi, ``A useful lemma for capacity analysis of fading interference
  channels,'' \emph{IEEE Trans. Commun.}, vol.~58, no.~2, pp. 411--416, Feb.
  2010.

\bibitem{6463498}
H.~Dhillon, R.~Ganti, and J.~Andrews, ``Load-aware modeling and analysis of
  heterogeneous cellular networks,'' \emph{IEEE Trans. Wireless Commun.},
  vol.~12, no.~4, pp. 1666--1677, Apr. 2013.

\bibitem{6575091}
D.~Cao, S.~Zhou, and Z.~Niu, ``Optimal combination of base station densities
  for energy-efficient two-tier heterogeneous cellular networks,'' \emph{IEEE
  Trans. Wireless Commun.}, vol.~12, no.~9, pp. 4350--4362, Sep. 2013.

\bibitem{6152217}
A.~Capone, A.~Fonseca~dos Santos, I.~Filippini, and B.~Gloss, ``Looking beyond
  green cellular networks,'' in \emph{Wireless On-demand Network Systems and
  Services (WONS), 2012 9th Annual Conf. on}, Jan. 2012, pp. 127--130.

\bibitem{6533307}
Y.~Chen, O.~Blume, A.~Gati, A.~Capone, C.-E. Wu, U.~Barth, T.~Marzetta,
  H.~Zhang, and S.~Xu, ``Energy saving: Scaling network energy efficiency
  faster than traffic growth,'' in \emph{Wireless Communications and Networking
  Conf. Workshops (WCNCW), 2013 IEEE}, Apr. 2013, pp. 12--17.

\bibitem{6918448}
A.~Shojaeifard, K.~Hamdi, E.~Alsusa, D.~So, and J.~Tang, ``A unified model for
  the design and analysis of spatially-correlated load-aware {HetNets},''
  \emph{IEEE Trans. Commun.}, vol.~62, no.~11, pp. 4110--4125, Nov. 2014.

\bibitem{6664198}
A.~Redondi, I.~Filippini, and A.~Capone, ``Context management in
  energy-efficient radio access networks,'' in \emph{Digital Communications -
  Green ICT (TIWDC), 2013 24th Tyrrhenian Int. Workshop on}, Sep. 2013, pp.
  1--5.

\bibitem{6497017}
Q.~Ye, B.~Rong, Y.~Chen, M.~Al-Shalash, C.~Caramanis, and J.~Andrews, ``User
  association for load balancing in heterogeneous cellular networks,''
  \emph{IEEE Trans. Wireless Commun.}, vol.~12, no.~6, pp. 2706--2716, Jun.
  2013.

\bibitem{6497439}
P.~Nardelli, M.~Kountouris, P.~Cardieri, and M.~Latva-aho, ``Throughput
  optimization in wireless networks under stability and packet loss
  constraints,'' \emph{IEEE Trans. Mobile Comput.}, vol.~13, no.~8, pp.
  1883--1895, Aug. 2014.

\bibitem{7478073}
A.~Shojaeifard, K.~A. Hamdi, E.~Alsusa, D.~K.~C. So, J.~Tang, and K.~K. Wong,
  ``Design, modeling, and performance analysis of multi-antenna heterogeneous
  cellular networks,'' \emph{IEEE Trans. Commun.}, vol.~64, no.~7, pp.
  3104--3118, July 2016.

\bibitem{6658810}
H.~Dhillon and J.~Andrews, ``Downlink rate distribution in heterogeneous
  cellular networks under generalized cell selection,'' \emph{IEEE Wireless
  Commun. Lett.}, vol.~3, no.~1, pp. 42--45, Feb. 2014.

\bibitem{6567878}
X.~Lin, J.~Andrews, and A.~Ghosh, ``Modeling, analysis and design for carrier
  aggregation in heterogeneous cellular networks,'' \emph{IEEE Trans. Commun.},
  vol.~61, no.~9, pp. 4002--4015, Sep. 2013.

\bibitem{6152071}
F.~Yilmaz and M.-S. Alouini, ``A unified {MGF}-based capacity analysis of
  diversity combiners over generalized fading channels,'' \emph{IEEE Trans.
  Commun.}, vol.~60, no.~3, pp. 862--875, Mar. 2012.

\bibitem{7110517}
J.~Tang, D.~K.~C. So, E.~Alsusa, K.~A. Hamdi, and A.~Shojaeifard, ``Resource
  allocation for energy efficiency optimization in heterogeneous networks,''
  \emph{IEEE J. Sel. Areas Commun.}, vol.~33, no.~10, pp. 2104--2117, Oct.
  2015.

\bibitem{opac-b1082765}
R.~P. Brent, \emph{Algorithms for minimization without derivatives}, ser.
  Prentice-Hall series in automatic computation.\hskip 1em plus 0.5em minus
  0.4em\relax Englewood Cliffs, N.J. Prentice-Hall, 1973.

\bibitem{NelderMead65}
J.~A. Nelder and R.~Mead, ``A simplex method for function minimization,''
  \emph{Computer Journal}, vol.~7, pp. 308--313, 1965.

\bibitem{ANAC:ANAC200410015}
S.~Singer and S.~Singer, ``Efficient implementation of the {Nelder–-Mead}
  search algorithm,'' \emph{Applied Numerical Analysis and Computational
  Mathematics}, vol.~1, no.~2, pp. 524--534, 2004.

\bibitem{Press:2007:NRE:1403886}
W.~H. Press, S.~A. Teukolsky, W.~T. Vetterling, and B.~P. Flannery,
  \emph{Numerical Recipes 3rd Edition: The Art of Scientific Computing},
  3rd~ed.\hskip 1em plus 0.5em minus 0.4em\relax New York, NY, USA: Cambridge
  University Press, 2007.

\bibitem{6619582}
F.~D. Cardoso, S.~Petersson, M.~Boldi, S.~Mizuta, G.~Dietl, R.~Torrea-Duran,
  C.~Desset, J.~Leinonen, and L.~M. Correia, ``Energy efficient transmission
  techniques for lte,'' \emph{IEEE Commun. Mag.}, vol.~51, no.~10, pp.
  182--190, Oct. 2013.

\bibitem{6056691}
G.~Auer, V.~Giannini, C.~Desset, I.~Godor, P.~Skillermark, M.~Olsson, M.~Imran,
  D.~Sabella, M.~Gonzalez, O.~Blume, and A.~Fehske, ``How much energy is needed
  to run a wireless network?'' \emph{IEEE Trans. Wireless Commun.}, vol.~18,
  no.~5, pp. 40--49, Oct. 2011.

\bibitem{6629714}
O.~Blume, A.~Ambrosy, M.~Wilhelm, and U.~Barth, ``Energy efficiency of {LTE}
  networks under traffic loads of 2020,'' in \emph{Wireless Communication
  Systems (ISWCS 2013), Proceedings of the Tenth Int. Symposium on}, Aug. 2013,
  pp. 1--5.

\end{thebibliography}

\appendices
\numberwithin{equation}{section}

\section{Aggregate Network Interference Statistics}
\label{BoundedMGFofInt}
The MGF of the aggregate network interference, considering a disk of radius $D$ around the reference user and then taking the limit as $D \rightarrow \infty$, can be derived as in 
\begin{align}
\mathcal{M}_{I_{agg,k} | R}(z) &  = \lim_{D \to +\infty} \mathbb{E}_{\varphi_{t,l,k},H_{t,l,k},\| \hat{Y}_{t,l,k} \|} \left\{ e^{-z \bigcup_{c \in \Phi^{(u)} / \{ k \}} \varphi_{t,l,c} \sum_{t \in \mathcal{T}} \sum_{l \in \Phi^{(b)}_{t} \backslash \{ l^{*}\} } P^{tx}_{t} H_{t,l,k} \chi_{t,l,k} \| \hat{Y}_{t,l,k} \|^{- \alpha_{t}}} \right\} \nonumber \\ & \overset{(a)}{\geq}  \lim_{D \to +\infty} \prod_{t \in \mathcal{T}} \prod_{l \in \Phi^{(b)}_{t} \backslash \{ l^{*} \}}  \mathbb{E}_{H_{t,l,k},\| \hat{Y}_{t,l,k} \|} \left\{ e^{ -  z  \overline{\varphi}_{t} P^{tx}_{t} H_{t,l,k} \| \hat{Y}_{t,l,k} \|^{- \alpha_{t}}}  \right\} \nonumber \\ & \overset{(b)}{=} \lim_{D \to +\infty} \prod_{t \in \mathcal{T}} \mathbb{E}_{\mathcal{N}^{(b)}_{t}} \left\{ \left( \mathbb{E}_{H_{t,j,k},\| \hat{Y}_{t,j,k} \|} \left\{ e^{-z \overline{\varphi}_{t} P^{tx}_{t} H_{t,j,k} \| \hat{Y}_{t,j,k} \|^{- \alpha_{t}}} \right\} \right)^{\mathcal{N}^{(b)}_{t}} \right\} \nonumber \\ & \overset{(c)}{=} \lim_{D \to +\infty} \prod_{t \in \mathcal{T}} \left( \rho_{t} \, \mathbb{E}_{H_{t,j,k},\| \hat{Y}_{t,j,k} \|} \left\{ e^{ -z \overline{\varphi}_{t} P^{tx}_{t} H_{t,j,k} \| \hat{Y}_{t,j,k} \|^{- \alpha_{t}} } \right\} + 1 - \rho_{t}  \right)^{\kappa_{t}} \nonumber \\ & \overset{(d)}{=} \prod_{t \in \mathcal{T}} e^{ -  \pi \lambda^{(b)}_{t,s} \mathbb{E}_{H_{t,j,k}} \left\{ \frac{\Gamma \left( 1 - \frac{2}{\alpha_{t}} \right) - \Gamma \left( 1 - \frac{2}{\alpha_{t}} , z  \overline{\varphi}_{t} P^{tx}_{t} H_{t,j,k} D^{- \alpha_{t}}_{t} \right)}{\left( z \overline{\varphi}_{t} P^{tx}_{t} H_{t,j,k} \right)^{\text{$\frac{- 2}{\alpha_{t}}$} }} + D^{2}_{t} \left( e^{ - z \overline{\varphi}_{t} P^{tx}_{t} H_{t,j,k} D^{- \alpha_{t}}_{t}} - 1 \right) \right\} }  \nonumber \\ & \overset{(e)}{=} e^{ - \pi \sum_{t \in \mathcal{T}} \lambda^{(b)}_{t,s} \left\{ \frac{ \Gamma \left(1- \frac{2}{\alpha_{t}} \right) \Gamma \left(m_{t}+ \frac{2}{\alpha_{t}} \right)}{ \Gamma (m_{t}) \left( \frac{z \overline{\varphi}_{t} P^{tx}_{t}}{m_{t}} \right)^{\text{$\frac{- 2}{\alpha_{t}}$}}} +  \frac{  m_{t}^{m_{t}} {\left( \frac{z \overline{\varphi}_{t} P^{tx}_{t} \beta_{t^{*}} P^{tx}_{t^{*}} }{\beta_{t} P^{tx}_{t} R^{ \alpha_{t^{*}} }}  +{m_{t}} \right)^{{- m_{t}}}}  - 1 }{\left( \frac{\beta_{t} P^{tx}_{t} R^{ \alpha_{t^{*}} }}{\beta_{t^{*}} P^{tx}_{t^{*}}} \right)^{\text{$\frac{- 2}{\alpha_{t}}$}}} - \frac{ _2F_1 \left({m_{t}}+1,{m_{t}}+ \frac{2}{\alpha_{t}} ;{m_{t}}+ \frac{2}{\alpha_{t}} +1;  \frac{-{m_{t}}\beta_{t} R^{ \alpha_{t^{*}} }}{z \overline{\varphi}_{t} {\beta_{t^{*}} P^{tx}_{t^{*}}}} \right)}{ \frac{\left( z \overline{\varphi}_{t} P^{tx}_{t} \right)^{{m_{t}}}}{m_{t}^{{m_{t}}+1} \left( {m_{t}} + \frac{2}{\alpha_{t}} \right)^{-1}} \left( \frac{\beta_{t} P^{tx}_{t} R^{ \alpha_{t^{*}} }}{\beta_{t^{*}} P^{tx}_{t^{*}}} \right)^{{- m_{t}} - \text{$\frac{2}{\alpha_{t}}$} }} \right\}}
\label{mainInterference}
\end{align}
where $(a)$ follows from applying Jensen's inequality to a convex function such that
\begin{align}
\overline{\varphi}_{t} = 1 - e^{\frac{- \lambda^{(u)} \, \overline{\varphi}_{t,j,c}}{\lambda^{(b)}_{t,s}}};
\end{align}
$(b)$ is from the independence of the tiers of BSs with $\mathcal{N}^{(b)}_{t}$ being the total number of potentially interfering tier-$t$ sources and $j$ being an arbitrary tier-$t$ source; $(c)$ is from utilizing the Binomial distribution such that $\mathcal{N}^{(b)}_{t} \sim \text{Binomial}\left(\kappa_{t},\rho_{t} \right)$; using the uniformly-distributed locations of the interferers
\begin{align}
\mathcal{P}_{\| \hat{Y}_{t,j,k} \|}(r) = 
\begin{cases}
\frac{2 r}{ D^{2} - D_{t}^{2}} & D_{t} < r < D\\
0 & \text{elsewhere}
\end{cases}
\label{Disss2}
\end{align}
with $D_{t} = \Big( \frac{\beta_{t} P^{tx}_{t}}{\beta_{t^{*}} P^{tx}_{t^{*}}} \Big)^{\frac{1}{\alpha_{t}}} R^{\frac{\alpha_{t^{*}} }{\alpha_{t}} }$ being the distance of closest tier-$t$ source, and
\begin{align}
\mathbb{E}_{r} \left\{ e^{ -z r^{- \alpha}} \right\} = \frac{ 2 z^{\frac{2}{\alpha}}}{\alpha \left(D^2-D_{t}^2 \right)} \left[ \Gamma \left(\frac{- 2}{ \alpha} ,D^{-\alpha } z\right)-\Gamma \left(\frac{- 2}{ \alpha} ,D_{t}^{-\alpha } z \right) \right],
\label{InteIdent}
\end{align}
$(d)$ can be derived by taking the limits as $D \rightarrow + \infty$, $\kappa_{t} \rightarrow + \infty$, $\rho_{t} \rightarrow 0$, and utilizing the Poisson limit theorem with $\frac{\kappa_{t} \rho_{t}}{\pi D^2} = \lambda^{(b)}_{t,s}$; 
finally, $(e)$ is obtained by taking the average with respect to the Gamma-distributed fading power gain of the arbitrary interferer using
\begin{align}
\mathbb{E}_{h} \left\{ h^{\frac{2}{\alpha}} \right\} = \frac{ \Gamma \left(m+\frac{2}{\beta }\right)}{m^{\frac{2}{\beta} } \Gamma (m)} 
\end{align}
and
\begin{align}
\mathbb{E}_{h} \left\{ \Gamma \left(1-\frac{2}{\alpha },\beta h \right) h^{\frac{2}{\alpha} } \right\} = \frac{ m^{m+1}}{\beta^{ m + \frac{2}{\alpha } } \left( m+\frac{2}{\alpha} \right)} \, _2F_1\left(m+1,m+\frac{2}{\alpha };m+\frac{2}{\alpha }+1;-\frac{m}{\beta }\right).
\end{align}

\section{Simplified Average Rate Expression}
\label{SimplifiedRateExpression}

Utilizing (\ref{MGF1}) and (\ref{MGF2}) in (\ref{mainER2}), considering $\eta = 0$ and $m = 1$, the bounded average rate expression in single-tier scenarios can be written in a double integral form as in (\ref{RR1}).
\begin{figure*}[!t]
\normalsize
\begin{align}
\tilde{\mathcal{R}}(.) & = \log_{2}(e) \int_0^{+ \infty } \int_0^{+ \infty } e^{-2 \pi  \lambda^{(b)}_{s} \left( \frac{R^{\alpha+2} \, _2F_1\left(1,1+ \frac{2}{\alpha} ;2+\frac{2}{\alpha};\frac{-R^{\alpha}}{ \overline{\varphi} z P^{tx}} \right)}{(\alpha +2) \overline{\varphi} z P^{tx}} +\frac{\pi  \csc \left(\text{$\frac{2 \pi }{\alpha }$} \right) (\overline{\varphi} z P^{tx})^{\frac{2}{\alpha} }}{\alpha } \right)} \frac{2 \pi  \lambda^{(b)}  R P^{tx}}{R^{\alpha }+ z P^{tx}} \diff z \diff R
\label{RR1}
\end{align}
\hrulefill
\end{figure*}
By employing a change of variables with $z = \frac{u^{\alpha}}{\overline{\varphi} P^{tx}}$ and hence converting from Cartesian to polar coordinates with $R = r \sin (t)$ and $u = r \cos (t)$, (\ref{RR1}) reduces to (\ref{RR2})
\begin{figure*}[!t]
\normalsize
\begin{align}
\tilde{\mathcal{R}}(.) & = \log_{2}(e) \int_0^{\frac{\pi}{2}} \int_0^{+ \infty } e^{ -2 \pi  \lambda^{(b)}_{s} r^2 \left( \frac{ \sin^{2}(t) \tan^{\alpha }(t) \, _2F_1\left(1,1+ \frac{2}{\alpha};2+\frac{2}{\alpha};- \tan^{\alpha}(t) \right)}{\alpha +2} + \frac{\pi  \csc \left(\frac{2 \pi }{\alpha }\right) \cos^2(t)}{\alpha } \right)} \frac{2 \pi  \alpha  \lambda^{(b)}_{s}  r \cos^{\alpha }(t) \tan (t) }{ \overline{\varphi} \sin^{\alpha }(t)+\cos^{\alpha }(t) } \diff r \diff t
\label{RR2}
\end{align}
\hrulefill
\end{figure*}
By using the following integral identity (where $x > 0$)
\begin{align}
\int_0^{+ \infty } r e^{-x r^2} \diff r = \frac{1}{2 x}
\end{align}
we arrive at the simplified average rate expression in (\ref{SRate0}). For the special case of $\alpha = 4$, with variable substitution $s = \tan^{2}(t)$ and some basic algebraic  manipulations, (\ref{SRate0}) can be further simplified to obtain (\ref{SRateLower}). \hfill $\blacksquare$

\section{Monotonicity Analysis of the Rate Function}
\label{MonotoneFunc}

Without loss of generality, consider a homogeneous DenseNet with Rayleigh fading for the intended and interference channels and path-loss exponent being equal to four. Recall that the average rate in nat/s/Hz of an arbitrary user in this case can be expressed by
\begin{align}
\mathcal{R}(.) = \int_0^{+ \infty } \frac{4 \diff s}{\left( 1 + \left[ 1 - e^{ - \frac{ \lambda^{(u)}}{ \lambda^{(b)} }} \right] s^2 \right) \left(2 s-2 \arctan(s)+\pi \right)}.
\end{align}
To investigate the behavior of the rate function with respect to the deployment density, we differentiate using basic substitution the inside of the above integral with respect to $\lambda^{(b)}$ as 
\begin{multline}
\frac{\diff}{\diff \lambda^{(b)}} \left( \frac{\diff R(.)}{\diff s} \right) = \frac{4}{\left(2 s-2 \arctan(s)+\pi \right)}  \left( \frac{\diff}{\diff \lambda^{(b)}} \left( 1 + \left[ 1 - e^{ - \frac{ \lambda^{(u)}}{ \lambda^{(b)} }} \right] s^2 \right)^{-1} \right) \\ = \frac{2 \lambda^{(u)} e^{- \frac{\lambda^{(u)}}{\lambda^{(b)}}} s^2}{{\lambda^{(b)}}^2 \left(1 + \left[ 1-e^{-\frac{\lambda^{(u)}}{\lambda^{(b)} }}\right] s^2 \right)^2 \left( s- \arctan(s)+\frac{\pi}{2} \right)} > 0
\label{diffDep} 
\end{multline}
where (\ref{diffDep}) follows given $\lambda^{(u)}$ only takes on positive values and $s- \arctan(s)+\frac{\pi}{2} > 0$ holds for positive values of $s$. This proves that the average rate is a strictly monotone function for BS deployment density. $\blacksquare$

\section{Closed-Form Average Rate Bounds}
\label{rateApprox}

Utilizing the following tight approximation 
\begin{align}
\arctan(s)  =  \arcsin \left( \sqrt{\frac{s^2}{1 + s^2}} \right)  \geq \frac{s}{1 + s} 
\end{align}
we can respectively obtain from (\ref{SRateLower}) and (\ref{SRateUpper}) 
\begin{align}
\tilde{\mathcal{R}}(.) = \log_{2}(e) \int_0^{+ \infty }  \frac{4 (1 + s)}{\left( 1 + \overline{\varphi} s^2 \right) \left(\pi ( 1 + s ) + 2 s^2 \right)} \diff s
\end{align}
and
\begin{align}
\tilde{\mathcal{R}}(.) = \log_{2}(e) \int_0^{+ \infty} \frac{2 (1 + s)}{\left( 1 + \overline{\varphi} s^{2} \right) \left( 1 - s + s^{2} \right)} \diff s. 
\end{align}
By performing a partial fraction decomposition we respectively obtain 
\begin{align}
\tilde{\mathcal{R}}(.) = \log_{2}(e) \frac{1}{1 + \overline{\varphi} \left( \pi \left( \frac{\pi}{4} \left( \overline{\varphi} +1 \right) - 1 \right)  \right)} \int_0^{+ \infty } \frac{2 s - \pi \overline{\varphi} + 2}{ \frac{\pi}{2} ( s + 1 ) + s^2 } + \frac{\overline{\varphi} (\pi ( \overline{\varphi} + 1)  - 2 ( s + 1) )}{1 + \overline{\varphi} s^2} \diff s.
\label{SRate33}
\end{align}
and
\begin{align}
\tilde{\mathcal{R}}(.) = \log_{2}(e) \frac{2}{1 - \overline{\varphi} + \overline{\varphi}^{2}} \int^{+ \infty}_{0} \frac{s - \overline{\varphi} + 1}{1 + s + s^2} + \frac{\overline{\varphi} ( \overline{\varphi} - s)}{1 + \overline{\varphi} s^2} \diff s
\end{align}
To continue, we present the following integral identities (where $\alpha > \frac{1}{4}$ and $\zeta \geq 0$)
\begin{align}
\int_0^{+ \infty } \frac{1}{ 1 + s + \alpha s^2} \diff s = \frac{\pi - 2 \arctan \left( \frac{1}{\sqrt{4 \alpha - 1}} \right)}{\sqrt{4 \alpha - 1}}
\label{id1}
\end{align}
and
\begin{align}
\int_0^{+ \infty } \frac{1}{1 + \zeta s^2} \diff s = \frac{\pi }{2 \sqrt{\zeta}}.
\label{id2}
\end{align}
From (\ref{id1}), (\ref{id2}), and using the rules of integration by parts, we arrive at the closed-form bounded expressions of the average rate in (\ref{LALB}) and (\ref{LAUB}). \hfill $\blacksquare$

\section{Optimal Deployment Density}
\label{OptSolBett}

The closed-form bounded expressions of the average rate in (\ref{LALB}) and (\ref{LAUB}) are complex highly non-linear functions of the ratio of the BS over the UE spatial densities. As a result, it is not possible to directly derive an expression for the optimal deployment solution. By taking the limits as $\lambda^{(u)} \rightarrow 0$ (sparse traffic) and $\lambda^{(u)} \rightarrow + \infty$ (full traffic), it can be readily deduced that $0 \leq \overline{\varphi} = 1 - \frac{\lambda^{(u)}}{\lambda^{(b)}} \leq 1$. Hence, we can apply the lower-bound approximation $\ln (1 + \overline{\varphi}) \leq \overline{\varphi}$ based on Taylor series expansion in (\ref{LALB}) and (\ref{LAUB}), in order to rearrange $\mathcal{R}\big(\lambda^{(u)},\lambda^{(b)},P^{tx},\eta,\alpha,m,\mu,\sigma \big) = \mathcal{R}_{0}$ and develop upper-bound and lower-bound closed-form approximations for the optimal network density ${\lambda^{(u)}}^{*}$ based on the real positive real roots of high-order polynomial functions in (\ref{optWC}) and (\ref{optBC}). \hfill $\blacksquare$

\section{Load-Aware Monte-Carlo Simulations}
\label{Monte-Carlotrials}

\begin{itemize}

\item[1)] Set the UEs density, and for each tier, select transmit power, BSs density, path-loss exponent, Nakagami-m fading, and Log-Normal shadowing mean and variance. 

\item[2)] Define a region of sufficiently large area around reference UE situated at the origin.

\item[3)] Generate the statistical numbers of tiers of BSs and UEs.

\item[4)] Deploy Uniformly-distributed heterogeneous BSs and UEs around the specified area.

\item[5)] Generate Nakagami-m fading and Log-Normal shadowing gains from all links.

\item[6)] Assign the reference UE to the BS which provides the strongest received shadowed power.      

\item[7)] Optimally and exclusively associate every other UE to a BS. Search through all BSs and if a BS is associated with one or more UEs it is active; otherwise is not transmitting.  

\item[8)] Compute the aggregate network interference experienced by the reference UE using the sum of received signal powers from only the interfering BSs. 

\item[9)] Calculate the reference UE SINR and evaluate the average rate.

\item[10)] Repeat steps (3) to (9) for a sufficiently large number of times and take the average.    
\end{itemize}


\end{document}